\documentclass[iop]{emulateapj}
\usepackage{epsfig}
\usepackage{color}
\usepackage{natbib, amsmath, amssymb}
\usepackage{apjfonts}
\begin{document}
\title{Kinematics of the Stellar Halo and the Mass Distribution of the Milky Way using BHB stars}
\author{PRAJWAL R. KAFLE$^{1}$, SANJIB SHARMA, GERAINT F. LEWIS \& JOSS BLAND-HAWTHORN }
\affil{Sydney Institute for Astronomy, School of Physics, A28, The University of Sydney, NSW 2006, Australia}
\email{$^{1}$ p.kafle@physics.usyd.edu.au}

\def \vl {v_{l}}                   \def \vb {v_{b}}
\def \vr {v_{r}}                   \def \vtheta {v_{\theta}}
\def \vphi {v_{\phi}}              \def \vlossigma {\sigma_{los}}
\def \rsigma {\sigma_{r}}          \def \tsigma {\sigma_{\theta}}
\def \psigma {\sigma_{\phi}}       \def \sigmalos {\sigma_{\rm los}}
\def \vrot {v_{\rm rot}}           \def \vlos {v_{\rm los}}
\def \vcirc {v_{\rm circ}}
\def \kpc {\rm kpc}
\def \msun {M_{\sun}}

\def \path {/home/atharv/Dropbox/papers_sync/BIB} 

\begin{abstract}

Here we present a kinematic study of the Galactic halo out to a radius of $\sim60$ kpc, using 
4664 blue horizontal branch (BHB) stars selected from the SDSS/SEGUE survey, to determine 
key dynamical properties. Using a maximum likelihood analysis, we determine 
the velocity dispersion profiles in spherical coordinates ($\rsigma$, $\tsigma$, $\psigma$) 
and the anisotropy profile ($\beta$). The radial velocity dispersion profile ($\rsigma$) is 
measured out to a galactocentric radius of $r \sim 60$ kpc,
but due to the lack of proper-motion information,  $\tsigma$, $\psigma$ and $\beta$ could only be 
derived directly out to $r \sim 25$ kpc.  From a starting value of $\beta\approx 0.5$ in the inner 
parts ($9<r/\kpc<12$), the profile falls sharply in the range $r \approx 13-18$ kpc, with a
minimum value of $\beta=-1.2$ at $r=17$ kpc, rising sharply at larger radius.
In the outer parts, in the range $25<r/\kpc<56$, we predict the
profile to be roughly constant with a value of $\beta\approx 0.5$.
The newly discovered kinematic anomalies are shown {\it not} to arise from
halo substructures. 
We also studied the anisotropy profile of simulated stellar halos formed purely
by accretion and found that they {\it cannot} reproduce the sharp dip seen in the data. 
From the Jeans equation, we compute the stellar rotation curve ($\vcirc$) of the Galaxy out to  
$r \sim 25$ kpc. 
The mass of the Galaxy within $r \lesssim 25$ kpc is determined to be
$2.1 \times 10^{11}$ $\msun$, and with a 3-component fit to $\vcirc(r)$, we determine the virial mass 
of the Milky Way dark matter halo to be $M_{\rm vir} = 0.9 ^{+0.4}_{-0.3} \times 10^{12}$ $\msun$ 
($R_{\rm vir} = 249^{+34}_{-31}$ kpc).
\end{abstract}

\section{INTRODUCTION}
Understanding the formation of stellar halos gives vital insight into 
the formation history and the evolution of galaxies 
\citep{1993ARA&A..31..575M, 2002ARA&A..40..487F, 2008A&ARv..15..145H}. 
Under the currently favored $\Lambda$CDM model of galaxy formation, 
the stellar halos are thought to have been built up, at least in part, 
 by accretion of satellite galaxies
\citep{1978MNRAS.183..341W, 1978ApJ...225..357S, 1999MNRAS.307..495H}.
Discoveries of structures like the Sagittarius dwarf galaxy 
\citep{1994Natur.370..194I, 1995MNRAS.277..781I, 2003ApJ...599.1082M, 2006ApJ...642L.137B}, 
the Virgo over-density \citep{2008ApJ...673..864J}, 
the Triangulum-Andromeda structure \citep{2004ApJ...615..732R, 2004ApJ...615..738M, 2007ApJ...668L.123M}
and the low-latitude Monoceros ring \citep{2002ApJ...569..245N}
lend further support to the idea of the stellar halo being formed by
accretion. Other than accretion, a part of the stellar halo could 
also be formed by in-situ stars.
Recent hydro-dynamical simulations \citep{2006MNRAS.365..747A, 2009ApJ...702.1058Z, 
2011MNRAS.417.1260F, 2012MNRAS.420.2245M} of galaxy formation suggest 
that in the inner regions the stellar halo might be dominated by in-situ stars, 
whose kinematic properties are distinct from the accreted stars. 

Observational evidences of multi-component halo have been found in 
dynamical studies of SDSS calibration stars \citep{2007Natur.450.1020C, 2010ApJ...712..692C, 2012ApJ...746...34B},
in rotational behavior in metallicity bins \citep{2011MNRAS.411.1480D},
in kinematics of Galactic anti-center and North Galactic Pole population \citep{2007MNRAS.375.1381K},
in chemical properties \citep{2010ApJ...714..663D},
and also in age difference between in-situ and accreted halo \citep{2012Natur.486...90K}.
There also exists a counter-claim by \cite{2011MNRAS.415.3807S}, who demonstrate that
the evidence of retrograde signal in the outer halo in \cite{2010ApJ...712..692C} 
sample is weak and is because of a manifestation of incorrect distance estimates. 
Investigating, whether any signal of multi-component halo could also be seen 
in the dispersions of the velocity of the halo population is important. 
Ultimately, studying the velocity dispersion profiles of the halos and comparing 
them with simulations might help to constrain the model of galaxy formation. 

Lack of proper motions and the slightly off centered position 
of the Sun with respect to the galactic center poses a unique challenge in studying the
kinematics of the stellar halo. 
At distances much larger than $R_{\sun}$, line of sight velocity is same as radial velocity 
with respect to the galactic center. 
Hence the $\rsigma$ profile is easy to compute at large distances and this has been well studied;
\cite{2005MNRAS.364..433B} studied the line-of-sight velocity dispersion
($\vlossigma$) of a mixture of 240 halo objects and found that
$\vlossigma$ decreases monotonically beyond $r$ $\sim$ 30 kpc.
In the outer most halo at $r\sim100$ kpc \cite{2005MNRAS.364..433B} and recently, \cite{2012MNRAS.425.2840D}
both see a significant drop in $\sigmalos$ value to $\sim50$ kms$^{-1}$.

Conversely, \cite{2010ApJ...719.1582D} studied 666 BHB stars from the 2QZ Redshift Survey
and found the velocity dispersion profile increases at large distances. 
But using 910 distant halo A-type stars, \cite{2010AJ....139...59B} 
found that there is a mean decline of ${\rm -}0.38$ $\pm$ 0.12 kms$^{-1}$ kpc$^{-1}$ in
$\rsigma$ over 15 $< r/\kpc < 75$. 
More recently, \cite{2008ApJ...684.1143X} used 2401 BHB halo stars 
within 60 kpc and measured a slower decline in $\vlossigma$ compared to earlier studies. 
At small $r$ it has been difficult to derive the $\rsigma$
profile, and the only attempt to measure the $\rsigma$ in inner-halo
was undertaken by \cite{1997ApJ...481..775S}. 
They find a sharp fall in $\rsigma$ from $\sim140$ kms$^{-1}$ to $\sim100$ kms$^{-1}$ 
at $r\approx12$ kpc although they assume the circular velocity to be constant. 

In a solar neighborhood one can get useful proper motions and with 
this all the three velocity dispersions 
($\rsigma$, $\tsigma$, $\psigma$) can be measured. \cite{2009ApJ...698.1110S} used the full phase space 
information of $\sim 1700$ halo subdwarfs from the solar neighborhood ($<5$ kpc)
and determined the velocity dispersions to be ($\rsigma$, $\tsigma$, 
$\psigma$) = ($143\pm2$, $82\pm2$, $77\pm2$) kms$^{-1}$. 
Also, \cite{2010ApJ...716....1B} analyzed the proper motions of 
a large sample of main-sequence stars within the solar neighborhood ($<10$ kpc) 
and found $\rsigma=141$ kms$^{-1}$, $\tsigma=75$ kms$^{-1}$, $\psigma=85$ kms$^{-1}$. 
A summary of the estimated values of the velocity dispersions found in the literature is 
given in Table \ref{table:dispersion_table}. 

\begin{table*}
\centering
\caption{Velocity Dispersions and Anisotropy Reference Table }
\begin{tabular}{ccccccccc}
\hline
\hline
Sample (number) & distance (kpc) & $\rsigma, \tsigma, \psigma$ (kms$^{-1}$) & anisotropy ($\beta$) & Reference\\
\hline
BHB ($\sim700$) & ($r\lesssim 20$, $r\gtrsim45$) & $\rsigma = (140, 90-110)$ & (radial(0.5),tangential(-1.3)) & \citealt{1997ApJ...481..775S}   \\
BHB (1170) & $5\lesssim d \lesssim 96$ & $101.4\pm2.8$, $97.7\pm16.4$, $107.4\pm16.6$ & nearly isotropic & \citealt{2004AJ....127..914S}  \\ 
BHB (1933) & $16<r<48$ & - & radial ($0.5^{+0.08}_{-0.2}$) & \citealt{2012MNRAS.tmpL.469D} \\
Subdwarfs ($\sim1700$)  & $d<5$ & $143\pm2$, $82\pm2$, $77\pm2$ & radial ($\sim0.69$) & \citealt{2009MNRAS.399.1223S}    \\
MS (10$^{5}$) & $d\backsimeq10$ & 141,75,85 & radial ($\sim0.67$) &  \citealt{2010ApJ...716....1B} \\
BHB (3549) & ($10<r<25$, $25<r<50$) & - & (tangential($-0.6$),radial($0.5$)) & \citealt{2011MNRAS.411.1480D}&\\
\hline
\end{tabular}
\label{table:dispersion_table} 
\end{table*}

Due to the lack of reliable proper motions of the halo field stars, 
even the fundamental quantities like the tangential components of the velocity dispersions as well as 
the anisotropy ($\beta$) are still badly understood beyond solar neighborhood.
The situation is however not as hopeless as it might seem.
At small $r$, using only line of sight velocity it is possible 
to put some constraints on these quantities using maximum likelihood techniques, 
where a model or a distribution function needs to be specified {\it a priori}. 
In an analysis of $1170$ BHB stars ranging from $5\lesssim d/\kpc \lesssim96$, where
$d$ is now the stellar distance rather than a radius,
\cite{2004AJ....127..914S} fit an ellipsoidal distribution of velocities and find that the halo is isotropic. 
Similarly, \citet[hereafter D11]{2011MNRAS.411.1480D} fit a constant
anisotropy model (power law Distribution Function) to 3549 BHB stars constructed 
from the SDSS Data Release 7 (DR7) and find that the halo between 
$r = 10-25$ kpc is tangential whereas the distant halo within $25<r/\kpc<50$ is radial. 
D11 claims for the tangential inner halo are in contrast with the result by
\cite{2010ApJ...716....1B} who found the inner halo to be radial in the similar 
region ($d\approx 10$ kpc). 
D11 assume the distribution function (DF) to be such that the tracer density and the potential 
both are power laws. 
Without {\it a priori} knowledge of the density slope their estimates will have some systematics. 
Additionally, the potential was also kept constant in their analysis 
and thus can bias the results due to the degeneracy between the potential and the anisotropy. 

In their most recent work, \citet[hereafter D12]{2012MNRAS.tmpL.469D} 
allow both the potential and $\beta$ to vary and thereby break the degeneracy,
and find that the outer halo within $16< r/\kpc<48$ is radial with $\beta =0.5^{+0.1}_{-0.2}$. 
Previous estimates of the velocity anisotropy ($\beta$) in the solar
neighborhood and the nearby halo are summarized in Table \ref{table:dispersion_table}. 
All the above estimates of an anisotropy of the distant halo
is done in a large radial bins. 
Their results might be accurate for the given radial bins and could also 
be the actual anisotropy of the halo provided the anisotropy
remains nearly constant or monotonic through out. On the other hand,
if the actual $\beta$ of the system is not monotonic 
but has a more complex shape, then estimating it in the larger 
bin will just capture an overall property of that bin. 

Theoretically there are families of the distribution function of \citet{1973A&A....24..229H} 
type which result in a constant anisotropic system, as well with the families of 
models those have their own anisotropy profiles, which include Osipkov-Merritt model
(\cite{1979SvAL....5...42O, 1985MNRAS.214P..25M, 1985AJ.....90.1027M}), 
\cite{1991MNRAS.250..812G}, \cite{1991MNRAS.253..414C}, \cite{2007A&A...471..419B}
and few more with the Hernquist potential-density model in \cite{2002A&A...393..485B}. 
The question we ask is how well do these anisotropy profiles match the 
anisotropy of the Galaxy? More fundamentally, how well do we know the
anisotropy of the halo? 
To this end we thus compute the beta profile with much finer spatial resolution and without 
any prior assumptions about density or potential. 

Another use of studying the kinematics of the stellar halo is to 
constrain the mass and the potential of the Milky Way. 
Most of the methods to estimate the mass require the knowledge of the 
anisotropy parameter $\beta$.  
Without the unbiased estimate of the velocity anisotropy, constraining the
mass of the system via the Jeans equation could be uncertain by
73\% (for $-4.5<\beta<0.44$) as found by \citealt{2010MNRAS.406..264W}
in their estimates of mass of the Galaxy. Several other authors have
also used this assumption to estimate the mass of  the galaxy
\citep{2006MNRAS.369.1688D, 2010ApJ...720L.108G,2012MNRAS.425.2840D}.
Precaution needs to be taken while making an arbitrary assumption about the anisotropy.
It has been found that the halo stars, the satellites and the dark matter halo have 
different orbital properties \citep{2006MNRAS.365..747A, 2007MNRAS.379.1464S}. 
Hence assuming a constant anisotropy for both field stars as well as satellites 
\citep{2005MNRAS.364..433B, 2006MNRAS.369.1688D} 
could introduce systematic uncertainties in the mass estimate. 
Ideally, in order to break the degeneracy we must have a separate estimation 
of the radial velocity dispersion, the velocity anisotropy
and underlying density of the population, as pointed out by \citet{2006MNRAS.369.1688D}.

The orbital evolution of the Magellanic clouds \citep{1982MNRAS.198..707L, 2007ApJ...668..949B}, 
the local escape speed \citep{2007MNRAS.379..755S}, the timing argument \citep{2008MNRAS.384.1459L}
and the study of the kinematic of the tracers population \citep{1996ApJ...457..228K, 
2008ApJ...684.1143X, 2010ApJ...720L.108G, 2010MNRAS.406..264W} are the methods undertaken 
in order to constrain the mass of the Galaxy. Summarizing all these attempts to 
constrain the mass of the Galaxy, the consensus can be found between $0.5-3.5\times10^{12}$ $\msun$.
Recently using BHB stars \cite{2008ApJ...684.1143X} estimate the 
mass of the Milky Way to be $0.91_{-0.18}^{+0.27} \times 10^{12} M_{\sun}$. However,
they make an assumption that the variation of $(\vcirc/\vlos)$, the ratio of the circular to the line
of sight velocity, with radius 
is same as that in simulations. 
In this paper, we estimate $\vcirc$ as far out as possible without any 
assumption and then use it to estimate the dark matter halo mass.

This work focuses mainly on the study of the kinematics of the stellar halo
in order to present the unbiased estimation of the velocity dispersions, 
anisotropy parameter and circular velocity as a function of radius to the extent data supports. 
We use a DF which does not require any assumption to be made 
{\it a priori} about the density profile or the potential. 
We then use our measurements of velocity dispersions to estimate the rotation curve of the Galaxy. 
The disc and bulge mass already being constrained from \cite{2009PASJ...61..227S}, 
we focus on constraining the dark matter halo mass. 
Using the circular velocity curve ($\vcirc(r)$) we can estimate $\beta(r)$ out 
to as far as $\rsigma(r)$ is available. 
Finally, we compare our results with simulations in which the halo 
is formed purely by accretion. 

This paper is organized as follows: in Section \ref{sec:theory} we discuss the 
theoretical aspect of our analysis, the methodology adopted, and the details 
about the sample; in Section \ref{sec:dispersion_profile} we present our 
result for the velocity dispersions, confer the results between the alternative models,
and investigate the contribution of the halo substructures ; in Section \ref{sec:circ_velocity} 
we present our estimation of the mass of the Galaxy. 
Results are then compared with the simulations in Section \ref{sec:simulated_halo}.
In Section \ref{sec:conclusion} we present our conclusion and discuss our result.
  
\section{THEORY AND METHOD}\label{sec:theory}

We are interested in calculating the velocity dispersions
$(\rsigma,\tsigma,\psigma)$ for the stellar halo. However, the data we
have is line of sight velocities. To proceed we need to make some 
assumptions about the position and the velocity of the Sun with respect to the 
galactic center. We assume ${\rm R}_{\sun}=-8.5$ kpc, the velocity of the 
local standard of rest (LSR), v$_{\rm LSR}$, is taken to be IAU 
adopted value = 220 kms$^{-1}$, and the solar motion with respect 
to LSR  (U, V, W)$_{\sun}$ = +11.1, +12.24, +7.25 in kms$^{-1}$ 
\citep{2010MNRAS.403.1829S}.
Spherical and heliocentric coordinate system are expressed in terms 
of ($r$, $\theta$, $\phi$), and ($d$, $l$, $b$) respectively.

\subsection{Distribution Function}
The distribution function (DF), f, is defined such that
f({\bf x},{\bf v}) d$^{3}${\bf x} d$^{3}${\bf v} 
is the probability of finding a randomly picked star 
in a phase-space volume d$^{3}${\bf x} d$^{3}${\bf v}. 
In general, we consider the stellar halo as an anisotropic 
spherical system. The anisotropy is defined as 
\begin{equation}\label{eqn:anisotropy}
 \beta = 1 - \frac{\tsigma^{2} + \psigma^{2}}{2\rsigma^{2}},
\end{equation}
$\rsigma,\tsigma$ and $\psigma$ being the
velocity dispersions in spherical coordinates, and it 
describes the orbital structure of the system. The values 
of this parameter range from --$\infty$ for purely circular
trajectories to 1 for purely radial orbits. 

Families of DFs that generate the collisionless anisotropic spherical systems 
with constant or varying velocity anisotropy can be found in 
detail in \cite{2008gady.book.....B}. One such distribution function with constant
anisotropy is 
\begin{equation}\label{eqn:constant_ansitropy}
 f(E,L) = f(E) L^{-2\beta}.
\end{equation}
Here $E=\Phi(r)-(v^{2}/2)$ is the relative energy per unit mass and
$L$ is the modulus of the angular momentum vector per unit mass.
Recently, a DF given by Equation (\ref{eqn:constant_ansitropy}) 
with the energy term from \cite{1997MNRAS.286..315E}:
\begin{equation}\label{eqn:Evansfe}
 f(E,L)\propto E^{(\beta(\gamma-2)/\gamma)+(\alpha/\gamma)-3/2} L^{-2\beta},
\end{equation}
was used by \citet{2011MNRAS.411.1480D} and \citet{2012MNRAS.tmpL.469D}
to study the rotation, anisotropy and mass of the 
Galactic halo. The parameters
$\alpha$ and $\gamma$ are the logarithmic slopes of the 
density ($\rho \propto r^{-\alpha}$) and 
potential ($\Phi \propto r^{-\gamma}$) respectively.
Hereafter, we refer to this function as D11 DF. 

If one is interested in deriving the dispersion profiles, a simple 
distribution function that one can use is the Gaussian velocity ellipsoidal
distribution function (GVE DF). 
The GVE DF has been used in the context of the stellar halo  by
\cite{1980MNRAS.193..295F} using globular clusters as tracers 
and by \cite{2004AJ....127..914S} and \cite{2009MNRAS.399.1223S} using halo stars.
A GVE DF with rotation about $z$-axis is given by

\begin{equation} \label{eqn:veDF}
f(r, {\bf v}) = \frac{\rho(r)}{(2\pi)^{3/2}\rsigma\tsigma\psigma} \text{exp}\left[-\frac{1}{2}\left(\frac{v_{r}^{2}}{\sigma_{r}^{2}}+
\frac{v_{\theta}^{2}}{\sigma_{\theta}^{2}}+\frac{(v_{\phi}-\vrot)^{2}}{\sigma_{\phi}^{2}}\right)\right]
\end{equation}

The DF as given by Equation (\ref{eqn:veDF}) assumes that the velocity
ellipsoid is perfectly aligned  with the spherical coordinates, but 
in general the velocity ellipsoid can have a tilt. 
Using halo subdwarf stars  \citet{2009ApJ...698.1110S} and  \citet{2010ApJ...716....1B}  
have found that the tilt is small and consistent with zero. Hence, 
it is safe to ignore the tilt while computing velocity dispersions. 

\subsection{Parameter estimation}
The proper motion information of the stars in the stellar halo beyond
solar neighborhood ($r\gtrsim10$ kpc) is not accurate enough to
properly constrain the tangential motions.
Nevertheless, our position in the Galaxy still makes it possible
to constrain these quantities by utilizing the tangential information carried by
the line-of-sight velocities of the stars. However, for that we need to marginalize
the distribution function over the unknown quantities, which in this
case are the tangential components ($v_{l}$, $v_{b}$).
The expression for the marginalized DF 
is given by,

\begin{equation}
\label{eqn:vlosd}
F(l, b, d, \vlos|\rsigma,\tsigma,\psigma,\vrot) = \iint f \text{d}v_{l} \text{d}v_{b}.
\end{equation}

We use maximum likelihood method to estimate the model parameters.  
The log-likelihood function which we maximize is given by
\begin{equation} 
\label{eqn:likelihood}
\mathcal{L} (l, b, d ,\vlos|\rsigma,\tsigma,\psigma,\vrot) =
\sum_{i}^{n}\text{log}F(l_{i}, b_{i}, d_{i}, v_{los_{i}}), 
\end{equation}
where $n$ is the number of stars in the system under study. 
We use Markov Chain Monte-Carlo (MCMC) with the Metropolis Hasting algorithm (MHA)  
to obtain the posterior distribution. We quote the central 
values of the velocity dispersions ($\rsigma, \tsigma, \psigma$) 
as our initial estimates and 16 and 84 percentiles as the error associated. 
Note, for the GVE DF the density term $\rho(r)$ in Equation (\ref{eqn:veDF}) is not
a function of model parameters and hence it does not
have any effect on the likelihood analysis.

Once the  radial velocity dispersion $\rsigma$ and the anisotropy parameter $\beta$ are
evaluated, Jeans equation \citep{1915MNRAS..76...70J} can be used to
estimate the circular velocity $v_{\rm circ}$ of the spherical system 
in equilibrium using the relation 
\begin{equation}\label{eqn:Jeanseq}
\vcirc^{2}(r) = -\rsigma^{2} \left[\frac{{\rm d\ln}\rho}{{\rm d\ln}
    r}+\frac{{\rm d\ln}\rsigma^{2}}{{\rm d\ln}r}+2\beta\right], 
\end{equation}
where $\rho \propto r^{-\alpha}$ is the density of the tracer population,
which implies ${\rm d} \ln \rho / {\rm d} \ln r = -\alpha$. 
Through out the analysis we assume the density to be double power law 
with $\alpha = 2.4$ ($r \leqslant 27$ kpc) and $\alpha = 4.5$ ($r > 27$ kpc)
in agreement with the recent works by \cite{2011MNRAS.416.2903D} and \cite{2009MNRAS.398.1757W}.

For systems with constant anisotropy and a given  $v_{\rm circ}$,  
the solution to the differential equation (\ref{eqn:Jeanseq}) 
subject to the boundary condition $\lim_{r \to \infty}$ $\rsigma^{2}$ = 0 reads 
\begin{equation}\label{eqn:Jeanseq_rsig}
 \rsigma^{2}(r) = \frac{1}{r^{2\beta}\rho(r)} \int_{r}^{\infty} dr' r'^{2\beta} \rho(r') ({\rm d}\Phi/{\rm d}r')
\end{equation} 
Assuming density and anisotropy are known  we can use this solution to 
estimate $\rsigma$ as a function of $r$.

\subsection{DATA: BHB \lowercase{stars}}\label{sec:data}

\begin{figure}
    \includegraphics[width=0.475\textwidth]{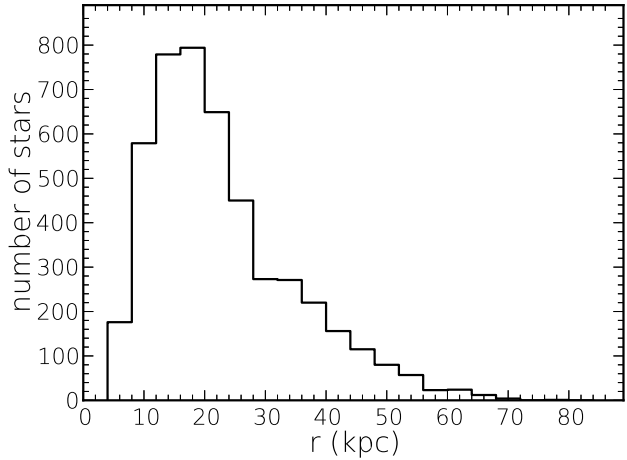}

  \caption{Radial distribution of BHB stars in galactocentric
    coordinates. The distribution has a peak at around 16 kpc. 
    Most of the stars are found to lie in the range $10<r/\kpc<25$.}
\label{fig:rgc}
\end{figure}

Being luminous and having nearly constant magnitude BHB stars are 
ideal for studying the stellar halo, and this is what we use 
in our study. We use the BHB catalog published by X11 for our analysis. 
The catalog comprises of 4985 BHB stars obtained from 
Sloan Digital Sky Survey (SDSS) Data Release 8
\citep{2011ApJS..193...29A}. The stars were selected 
by imposing limits on color and Balmer line profile
measurements. Imposing limits on Balmer line profile measurements 
allow one to remove the main sequence stars and Blue Stragglers. 
Further details on BHB candidate selection can be found in 
\citet{2008ApJ...684.1143X} and references therein.
To avoid contamination from the disk stars, we restrict our 
analysis to stars having a distance $|z|>4$ kpc from the galactic mid-plane. 
As mentioned earlier, the way our likelihood function 
(Equation \ref{eqn:vlosd}) is laid out, this cut in distance above the plane will not introduce 
any bias. No velocity limits have been imposed to obtain the sample and 
thus for the purpose of kinematic studies the population of BHB
stars we select can be considered to be kinematically 
unbiased.
 
For the stars that we study the angular position is known very
accurately, but the distance and radial velocity have some uncertainty 
associated with them. 
To get more accurate distances we recalibrate
X11 distances using a color-magnitude relation derived for the same population 
from \citet{2011MNRAS.416.2903D}. 
The estimated dispersion in g-band magnitudes is 0.13, equivalent to a distance uncertainty of 6$\%$. 
For the SEGUE \citep{2009AJ....137.4377Y} radial velocity measurements,  
94\% of our sample have an uncertainty of less than 8 kms$^{-1}$. 

The galactocentric radial distribution of the final BHB samples is
shown in Figure \ref{fig:rgc}. It can be seen that the distribution 
peaks at around 16 kpc. Most of the stars are found to 
lie in the range $10<r/\kpc<25$.

\section{Velocity Dispersion profile of the halo}\label{sec:dispersion_profile}

We study the kinematics of the halo in radial bins to obtain radial profile of 
the model parameters ($\rsigma, \tsigma, \psigma$) and also $\beta$. 
Using only line-of-sight velocity information the tangential 
components, $\tsigma$ and $\psigma$, are difficult to constrain 
except in the very inner regions of the halo. 
However, $\rsigma$ can be well constrained both in the inner and the 
outer halo. 
This means that relatively larger number of stars ($>$1000)
per bin are required to estimate $\tsigma$ and $\psigma$ as compared to $\rsigma$. 
Given that we only have about 4000 stars this means 
that we cannot measure the $\tsigma$ and $\psigma$ profiles with sufficient 
spatial resolution. 
Hence, we employ two different binning schemes or estimators, one for radial velocity and the other for 
tangential velocity. 
The estimators are; the equi-populated estimator (hereafter EPE) and central moving estimator (CME). 
In EPE the data is binned radially with each bin containing equal 
number of particles and this is used for computing $\rsigma(r)$. 
In CME a set of equi-spaced positions in $r$ are
chosen and then at each position an equal number of points either 
side of the chosen central value are used to estimate the desired quantity. 
We use the CME for computing $\tsigma(r)$ and $\psigma(r)$.
The crucial difference between the two schemes is that 
while the bins are non-overlapping in the former, in the latter they 
can be overlapping. 
In EPE the spacing between the bins is directly proportional to the 
number of particles in each bin $n_{\rm bin}$. Hence, if the desired
quantity can be estimated with sufficient accuracy employing small $n_{\rm bin}$,
then EPE is the desired method. However, if this is not the case then 
it is better to use the CME method as the spatial resolution is
not directly dependent on $n_{\rm bin}$
\footnote{For the effect of the bin size, see appendix \ref{sec:binning}.}. 

Finally, for our data the number density of points in $r$ is highly
non-uniform, and hence it is not accurate to assume that 
the desired quantity has been estimated at the center of the bin. 
To alleviate this number density bias, for both schemes we compute  
the final position of the bin as the mean $r$ of the points in the
bin.

\subsection{Radial velocity dispersion profile ($\rsigma$(r))}
\label{subsec:radialdis}
\begin{figure}
    \includegraphics[width=0.475\textwidth]{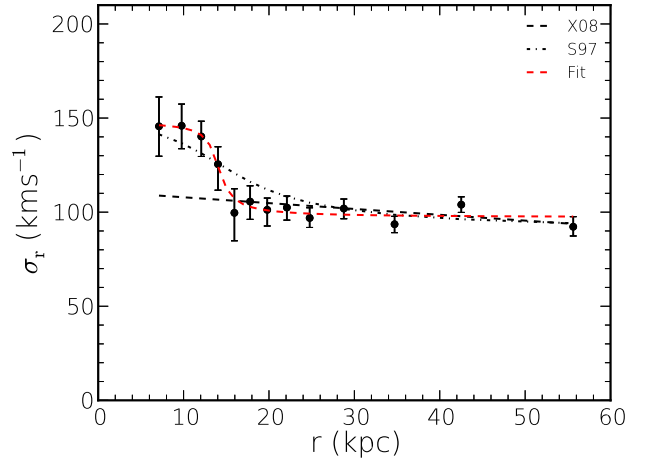}
  \caption{Radial velocity dispersion in radial bins. 
           The black dashed line is the $\sigmalos$ profile from
           \cite{2008ApJ...684.1143X}. Black dashed-dotted and red dashed lines
           are the Sommer-Larsen profiles given by Equation (\ref{eqn:rprofile}) 
           for the fitting parameters taken from \cite{1997ApJ...481..775S} and
           from the fit to our estimated values of $\rsigma$ respectively.}
\label{fig:sigmar_profile}
\end{figure}

Here we focus on the nature of the $\rsigma(r)$
profile of the halo for which we adopt the EPE method with $n_{\rm bin} = 400$. 
As explained previously, $\rsigma$ can be measured out to the extent of the data
($r\sim60$ kpc). 
The values of $\rsigma$ obtained from the likelihood analysis are 
given in Figure \ref{fig:sigmar_profile} and the error bars represent the 
1$\sigma$ confidence interval and are determined from the likelihood fitting. 

We find that the radial velocity dispersion, $\rsigma$ at the Sun's position 
(R$_{\sun}$ = 8.5) is 145.6 kms$^{-1}$. However, beyond the solar neighborhood
$\rsigma$ sharply decreases until $r\sim15$ kpc, after which it 
decreases much slowly and  approaches a value of around $\sim$100
kms$^{-1}$ at 56 kpc. 
The error bar in $\rsigma$ for the inner halo population is large and is 
mainly because the $\vlos$ contains less radial velocity 
information as compared to outer parts.
The overlaid black dashed line is the linear approximation
($\approx 111-0.31r$) for $\sigmalos$ profile from \cite{2008ApJ...684.1143X}.
Additionally, other previous attempts to fit the profile for $\rsigma$ 
in the outer parts of the halo \citep{2005MNRAS.364..433B, 2010ApJ...720L.108G, 2010AJ....139...59B}
also found profiles similar to \cite{2008ApJ...684.1143X},  
with slightly varying slope and normalization. 
All these profiles are reasonable estimates of $\rsigma$ for the outer 
halo ($d\gg$R$_{\sun}$) where the assumption $\rsigma \approx
\sigmalos$ holds. In the inner halo ($r\lesssim 15$ kpc) however the approximation 
breaks down and $\rsigma$ strongly deviates from $\sigmalos$. 
It can be seen that the deviation of $\rsigma$ from $\sigmalos$ increases as one
approaches the center and at R$_{\sun}$ the deviation is as high 
as $\sim 40$ kms$^{-1}$. 

\cite{1997ApJ...481..775S} provide a functional form for fitting 
the $\rsigma$ profiles which is given by 
\begin{equation}\label{eqn:rprofile}
 \rsigma^{2} = \sigma_{0}^{2} + \frac{\sigma_{+}^2}{\pi}\left[\frac{\pi}{2}-{\rm tan}^{-1} \left(\frac{r-r_{0}}{l}\right)\right].
\end{equation}
This has a shape which is similar to our $\rsigma(r)$ profile 
and we fit it to find $\sigma_0=94.5$ kms$^{-1}$, 
$\sigma_+=122.3$ kms$^{-1}$, $r_0=13.2$ kpc, and $l=2.6$ kpc.
In this function the fit parameter $\sigma_{0}$ gives the asymptotic value that
$\rsigma$ achieves in the outer halo, whereas ($\sigma_0^2 + \sigma_+^2)^{1/2}$
gives the approximate value for $\rsigma$ in the inner halo. 
The fit parameters $r_0$ and $l$ determine the turn-off point and  
the steepness of the transition of the profile respectively. A lower value of $l$
gives a steeper transition, this can be seen from the comparison between
the Sommer-Larsen fit ($l= 7.5$ kpc) and the red line  in Figure
 \ref{fig:sigmar_profile} which is our fit having smaller $l$. 

\subsection{Tangential velocity dispersion and anisotropy profiles ($\tsigma(r)$, $\psigma(r)$ and $\beta(r)$)}
\label{subsec:tangentialdis}

\begin{figure}
    \includegraphics[width=0.475\textwidth]{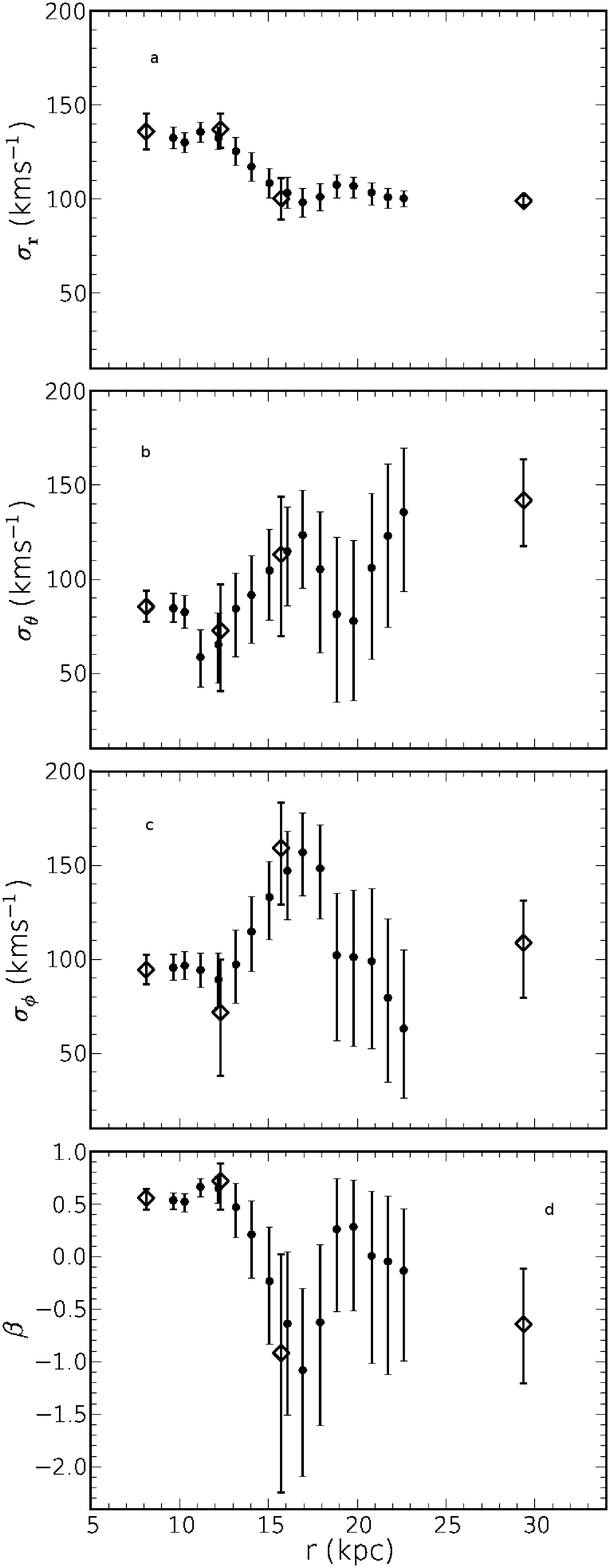}
   \caption{Velocity Dispersions and Anisotropy profiles: From top to bottom is $\rsigma,
   \tsigma, \psigma$, and $\beta$ profiles of the stellar halo estimated in the radial bins.
    The diamond and the round markers are the results for the two binning schemes
    , namely the EPE and the CME respectively. Note that the last radial bin marked with the open diamond contains 
    the remaining stars. The diamond markers in plot (a) is just shown for the ease of comparison.
            }
\label{fig:aniso}
\end{figure}

There have been few attempts to constrain the tangential velocity dispersions
and most of the studies are either restricted to solar neighborhood 
\citep{2009MNRAS.399.1223S, 2010ApJ...716....1B} or 
for the overall system  \citep{2004AJ....127..899S}.  
The tangential velocity dispersions not only provide the information of the anisotropy 
of the stellar population (through Equation \ref{eqn:anisotropy}) but together
with $\rsigma$ also helps to measure the mass distribution of the Galaxy.

We estimate the tangential velocity dispersions ($\tsigma$ and $\psigma$) 
using CME with $n_{\rm bin} = 1200$ stars. Given the quality of the data, 
we are only able to measure $\tsigma$ and 
$\psigma$ out to $\sim 25$ kpc. Our estimates of $\tsigma$ and 
$\psigma$ are shown in Figure \ref{fig:aniso}{b,c} by the black dots with error bars.
For uniformity, we also estimate the $\rsigma$ with this binning
scheme and this is
shown (only out to $\sim 25$ kpc) in Figure \ref{fig:aniso}{a} by the black dots 
with error bars.  

In general the tangential components $\tsigma$ and $\psigma$  near the solar neighborhood are found to be
comparatively lower $(\tsigma = 85_{-9}^{+8}$ kms$^{-1}$, $\psigma = 95_{-8}^{+8}$ kms$^{-1}$)
than the radial dispersion $\rsigma$. 
It can be seen in the Figure \ref{fig:aniso}{b and c} that there is 
a sharp rise in the values of $\tsigma$ and $\psigma$ at $r=17$ kpc. 
Beyond this $\psigma$  falls whereas $\tsigma$  rises, 
given the large uncertainties and low number of independent bins it is unclear if 
this is a real or a spurious trend.

By substituting the estimates of the tangential and the radial velocity 
dispersions obtained using CME $(n_{\rm bin} = 1200)$ from the above analysis 
into Equation (\ref{eqn:anisotropy}) we compute the corresponding values of the 
anisotropy constant in the respective bins. As shown in Figure 
\ref{fig:aniso}{d}, the halo within 12 kpc has $\beta \sim 0.5$ 
whereas the outer halo beyond the turn-off point is nearly isotropic.
We discover a significant drop in the $\beta$ profile at $r=17$ kpc.
Here the halo is strongly tangential with $\beta = -1.2$.
We later confirm that the trend observed in anisotropy is neither due to the manifestation of 
the systematics introduced by the chosen model ( \S \ref{sec:alternative_model}) 
nor due to presence of the halo substructures (\S \ref{sec:halo_substructures}). 
It is also shown in the appendix \ref{sec:vlsrrsun_effect}
that assumed v$_{\rm LSR}$ and R$_\sun$ have negligible effects upon 
these estimates.
The probable reasons for this sudden turn-over in the properties 
of the halo are discussed in the conclusion. 

We know that the consecutive CME bins overlap in radius
and thus the dispersion profiles demonstrated in Figure
\ref{fig:aniso} is a smoothed version of the actual dispersion profiles
of the halo. 
However, to check for any systematic associated 
with the choice of the binning scheme, we also  estimate  
$\rsigma, \tsigma, \psigma$, and  $\beta$ in traditional 
equi-populated (EPE) bins  $(n_{\rm bin} = 700)$. 
The measured values in these bins are shown by the diamond points in Figure \ref{fig:aniso}.
If the number of stars per bin is less than 700  it is difficult to 
constrain $\tsigma$ and $\psigma$. 
Even with $n_{\rm bin} = 700$ we were able to constrain the tangential motion only till 
$16$ kpc (first three diamond points). 
Hence, we construct the last bin by grouping all the stars beyond $16$ kpc into one bin 
(rightmost diamond point). 
More importantly, with this binning scheme we are only interested to see 
whether we obtain the corresponding dip or rise (depending on the parameter of interest) 
seen in Figure \ref{fig:aniso} or not.
We find that except for the right-most diamond points all of other
diamond points in the figure are in agreement with our previous estimates of
the dispersions (given by black dots). 
The rightmost diamond points are calculated in a huge bin with more than 50\% of the total sample.
Particularly for $\tsigma$, $\psigma$ and $\beta$ given the non-monotonic trend they have,
hence we do not claim the last diamond point is the correct estimate of anisotropy at 
$r\sim35$ kpc.    

None of the uncertainties quoted in the above estimates of $\rsigma$, $\tsigma$ and $\psigma$
include the uncertainties in distances and radial velocities. As mentioned earlier, the 
errors in distance and radial velocity are quite small, and convolving the  
model (Equation \ref{eqn:veDF}) with the error functions should not change the 
results. 
   
\subsection{$\beta(r)$ from fitting Distribution Function (f(E,L))}
\label{sec:alternative_model}
It would be interesting to see whether the $\beta$ profile
presented above, in particular the dip seen at $r=17$ kpc,
is an artifact of our chosen model (GVE) or
a real inherent feature of the Galactic halo. 
In order to pursue it, here we explore the effect of the chosen model 
on the determination of the velocity anisotropy ($\beta$). 
For the comparative study, the alternative model we choose is the D11 DF
(Equation \ref{eqn:Evansfe}).
We consider anisotropy ($\beta$) and potential $(= \Phi_{0}r^{-\gamma})$ as free model parameters
and constrain them using the maximum likelihood method.

\begin{figure*}
    \includegraphics[width=0.95\textwidth]{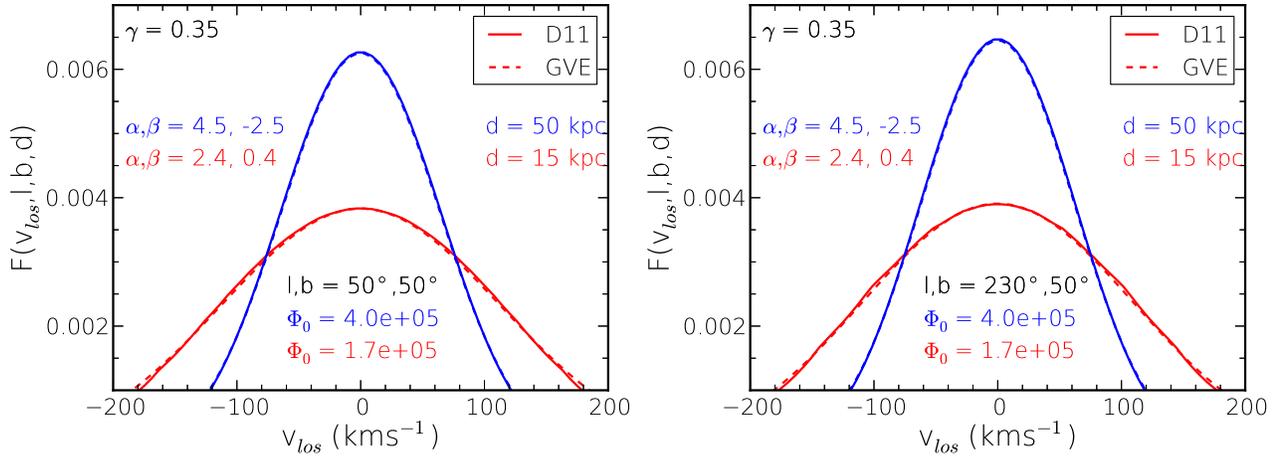}
  \caption{Comparison of the marginalized DF given by Equation (\ref{eqn:vlosd})
           obtained for D11 models and GVE distribution along
           two different line-of-sights. Solid line is the LOSVD for the D11 DF whereas
           dashed line is the LOSVD for the GVE model. Red and blue 
           lines represent LOSVD for two different distances (d) 15 kpc and 50 kpc
           respectively. The potential is assumed to be a power law given by 
           $\Phi_{\circ} r^{-\gamma}$, where $\Phi_{\circ}$ (in km$^{2}$s$^{-2}$) is the potential normalization.
           Potential slope ($\gamma$) is assumed constant and equal to
           0.35 for all the cases. Left: LOSVD along ($l$,$b$) = (50$^{\circ}$, 50$^{\circ}$)
           Right: LOSVD along ($l$,$b$) = (230$^{\circ}$, 50$^{\circ}$).}
\label{fig:VLOSD}
\end{figure*}

First, we compare the theoretical properties of the
DFs at our disposal namely, GVE and D11 DF. In Figure \ref{fig:VLOSD} we plot 
the theoretical LOSVDs of models along two separate line-of-sights. We also show the LOSVDs in Figure 
\ref{fig:VLOSD} for two different distances representing the inner halo 
($d$ = 15 kpc, given by red lines) and the outer halo ($d$ = 50 kpc, given by blue lines). 
For the inner-halo we assume $\alpha$ = 2.4 and assign radially biased 
anisotropy ($\beta$ = 0.4) whereas for the outer-halo we assume $\alpha$ = 4.5 and
assign tangentially biased anisotropy ($\beta$ = -2.5)
in accordance to \cite{2009MNRAS.398.1757W} and \cite{2011MNRAS.416.2903D}
estimates for $\alpha$. Note, the density normalization at the break radius ($r = 27$ kpc)
is assumed to be equal.
For the assumed constant potential the solid 
lines in Figure \ref{fig:VLOSD} (both left, right panels) are the LOSVDs 
obtained by adopting the D11 model and dashed lines are the LOSVDs of
our GVE model. 
Recall that our model does not take $\beta$ directly but demands the information 
of the velocity dispersion components ($\rsigma, \tsigma, \psigma$) individually. 
To make the LOSVDs obtained from both the models comparable we estimate 
$\rsigma, \tsigma, \psigma$ from the set of values of $\beta, \alpha, 
\Phi_{\circ}$ and $\gamma$ chosen to obtain LOSVDs of D11 DF. 
For an assumed potential power law we use Equation (\ref{eqn:Jeanseq_rsig})
to first calculate $\rsigma$, to put in our model. 
For an assumed $\beta$, substituting this $\rsigma$ in Equation (\ref{eqn:anisotropy}) gives the 
corresponding value for $\sigma_{\rm t}$ (= $\sqrt{\tsigma^{2} + \psigma^{2}}$). 
It can be seen in the figure that for all the four cases LOSVDs obtained 
from both the models match well. 
Naively, from these perfect matches of the LOSVDs at different 
line-of-sights and distances it can be anticipated that the estimation of 
$\beta(r)$ with both models should also match. 

\begin{figure}
    \includegraphics[width=0.475\textwidth]{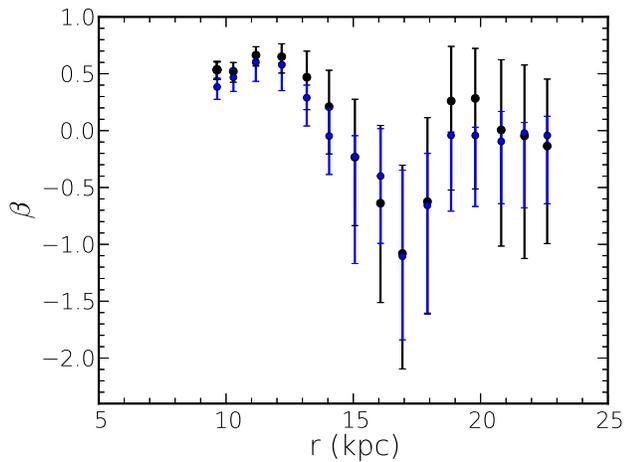}
   \caption{Anisotropy ($\beta$) estimates in the CME 
            using D11 model (blue points) and GVE model 
           (black points). Each bin consists of 1200 stars. Anisotropy 
            estimates with GVE distribution is done with 
            MCMC technique whereas estimates with D11 model is done with 
            the brute-force grid based analysis. Assigned asymmetric 
            uncertainties are 1$\sigma$ confidence intervals
            obtained from likelihood fitting.}
\label{fig:beta_2models}
\end{figure}

Now we estimate the $\beta(r)$ using exactly the same sample of BHB stars in same radial bins
as in \S \ref{subsec:tangentialdis} (CME, $n_{\rm bin} = 1200$) but with D11 DF. 
Blue points in Figure \ref{fig:beta_2models} 
demonstrate the $\beta$ profile obtained by fitting D11 DF. 
Here we use brute-force grid based analysis to constrain the model parameters  
$\beta$, $\Phi_{\circ}$ and  $\gamma$. 
To facilitate the comparison our estimates from Figure \ref{fig:aniso}{d}
are over plotted in Figure \ref{fig:beta_2models} and are shown by the black
points. From Figure \ref{fig:beta_2models} it can be seen that within the 
range of uncertainties the measured values of $\beta$ with both models 
(D11 DF and GVE model) agree. However, a slight bias can be seen in the 
sense that $\beta$ obtained from GVE DF is in general higher.
The reason for this discrepancy lies in the fact that $\beta$ obtained from 
the D11 DF has a dependence on $\alpha$. 
Hence, unless the underlying $\alpha$ value of the sample 
is exactly known, a mismatch is expected. 
The estimated value of $\beta$ increases with the adopted value of $\alpha$ (see Fig-3 D11).
This suggests that the actual value of $\alpha$ is even higher than the 
one that is adopted here (2.4).

\begin{figure}
    \includegraphics[width=0.475\textwidth]{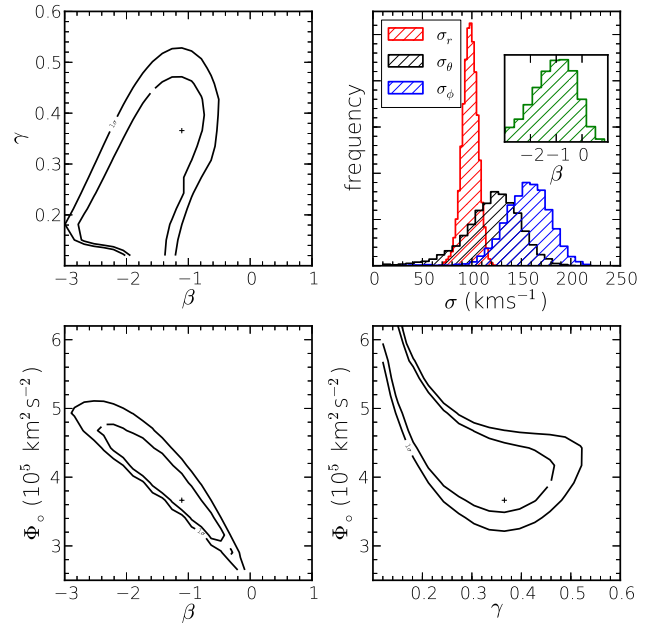}
  \caption{The posterior distributions of the parameters for bin centered at 
           r = 16.93 kpc. Upper Right: The posterior distributions of the 
           GVE model parameters, 
           $\rsigma$, $\tsigma$ and $\psigma$, obtained with 
           5$\times$10$^{5}$ MCMC random walks. In the inset is the derived 
           distribution of the $\beta$ parameter. Upper Left and Lower 
           panels: The joint likelihood contours of the D11 model
           parameters $\beta$, potential normalization ($\Phi_{\circ}$) and 
           potential slope ($\gamma$) are obtained with brute force analysis.
           The outer contour displays 1 $\sigma$ region whereas 
           the inner contour demonstrates a region of $50\%$ confidence interval.
           Cross hair corresponds to the point where the likelihood is maximum.}
\label{fig:r16.93_D11DF_&_veDF}
\end{figure}

In order to give an estimate of the quality of constraints obtained from 
the likelihood analysis we display the likelihood contours in the 
parameter space in the top left and bottom two plots in 
Figure \ref{fig:r16.93_D11DF_&_veDF} for a bin centered at 
$r$ = 16.93 kpc where the maximum dip in the $\beta$ profile was seen in 
Figure \ref{fig:beta_2models}. 
Additionally, in the corresponding bin, the top right plot in figure 
demonstrates the posterior distributions of the model parameters 
$\rsigma$, $\tsigma$ and $\psigma$ of the GVE 
model obtained from $5 \times 10^5$ MCMC random walks. 
It can be seen that even at a distance of just twice of R$_{\sun}$, 
$\tsigma$ and $\psigma$ distributions are quite broad as compared to 
$\rsigma$ distribution; this is the reason for the large 
uncertainty in the value of $\beta$ as we move outwards in $r$.

\subsection{Effect of the halo substructures}\label{sec:halo_substructures}
There is now enough observational evidence to suggest that the stellar
halo is highly structured particularly as one moves outwards into the
halo \citep{2008ApJ...680..295B}. 
Using clustering algorithms on simulated 
N-body stellar halos, \cite{2011ApJ...728..106S} find that the 
fraction of material in substructures increases monotonically
as a function of distance from the center and at around 
$65$ kpc can be as high $50\%$. 
Hence, while studying the kinematic properties of the halo  
should we include the substructures or exclude them? 
If the kinematic properties of a sample are dominated by 
a few massive accretion events then one should exclude them. 
However, in-spite of being highly structured if the sample 
is a superposition of large number of events with none 
of them being individually too dominant then it is best to include 
them. 
Results of \cite{2011ApJ...728..106S} on simulated halos show 
that for the range of radii that spans our BHB stars ($r<40 \ \kpc$) 
the amount of material in substructure should be less than 20\%. 
So we expect the substructures to have a minor effect on the kinematic 
properties that we have derived. However, it is still important 
to check if this is true. 

To study the effect of substructures on our estimation of the dispersions 
we mask two prominent features of the halo that contaminate our sample, 
namely, the Sagittarius stellar stream and the Virgo over-density. 
Cuts we impose include the Lambert Equal-Area projection cut as given in 
\citet{2008ApJ...680..295B} and an additional cut in equatorial coordinates (RA and DEC)
suggested by \citet{2011MNRAS.416.2903D}. We mask the region within $0<$ X 
(abscissa of the equal-area projection)$<30$, where X is given by 63.63961[2(1--sin $b$)]$^{1/2}$;
and $0^{\circ}< {\rm RA} <50^{\circ}$, and $-30^{\circ}< {\rm DEC} <0^{\circ}$,
which is purely a geometric cut. These stringent cuts reduce our final sample to 2975 stars.
A proper phase-space masking of these structures will be revised in the future work.   

\begin{figure}
    \includegraphics[width=0.475\textwidth]{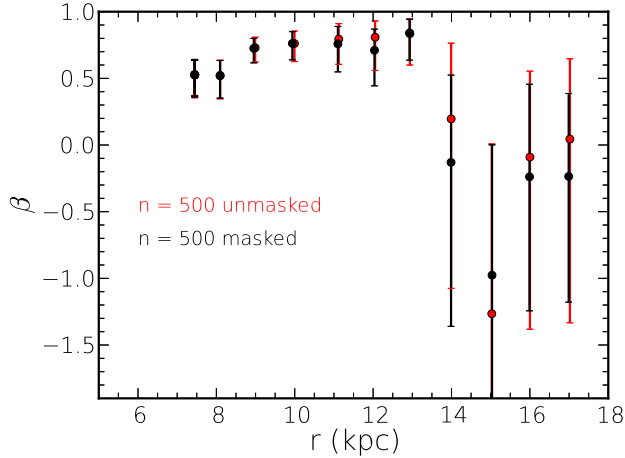}
  \caption{Effect of the halo substructures, namely, the Sagittarius stellar stream and the Virgo over-density. 
           Each CME bins contain 500 stars. Black and red points are our result for the 
           masked and the unmasked halo respectively. 
           Error bars quoted are 1$\sigma$ credibility interval.}
\label{fig:saggi_virgo}
\end{figure}

In Figure \ref{fig:saggi_virgo} we present our result obtained after
masking the substructures.
As masking reduces the sample size almost by half, 
we employ CME with $n_{\rm bin} = 500$ stars only, instead of $n_{\rm bin} = 1200$ as was done
earlier with unmasked data, to avoid excessive smoothing of the estimated profiles. 
Figure \ref{fig:saggi_virgo} shows that the masking of substructures 
have little effect on the estimation of velocity dispersion profiles,  
the $\beta$ profile is almost unchanged.
This alleviates the concern that perhaps the turn-over points in the velocity dispersion profiles
and the dip in $\beta$ profile, seen in Figure \ref{fig:aniso},
in the region $r=13-18$ kpc could be due to the dominance of halo substructures. 

\section{Comparison of anisotropy estimates in D11 and D12 radial bins}

In their recent work D11 and D12 fit a  distribution function 
of the form given by Equation (\ref{eqn:Evansfe}), to the BHB samples 
obtained from SEGUE survey in order to estimate the model parameters. 
The models adopted are constant anisotropy models given by 
Equation (\ref{eqn:constant_ansitropy}). 
In D11, the potential is assumed to be a power law ($\propto r^{-\gamma}$) and 
with a constant index ($\gamma = 0.35$). Later in D12, they break the degeneracy 
present in their model and consider the potential normalization ($\Phi_{0}$), 
potential slope ($\gamma$) and anisotropy ($\beta$) as free parameters. 
Note, in D11 there is an additional parameter specifying rotation 
(odd part of the DF) but this was dropped in D12, as they were not 
focusing on rotation.
The methodology applied to measure the model parameters is similar to ours
which involves marginalizing the DF over the tangential 
velocities to derive the line-of-sight velocity distribution (LOSVD);
fitting the LOSVD to the data using the maximum likelihood method and in return 
obtaining the best estimates of the model parameters. 

\subsection{Anisotropies of inner and outer halo by D11}\label{sec:D11comp}

D11 study the rotation and the anisotropy of the BHB samples taken from SDSS 
Data Release 7 \citep{2009ApJS..182..543A}. First, in order to construct the sample used by them we query  
SDSS DR7 database to 
select the candidate BHB stars using the color and 
the stellar parameters ranges given in D11. 
Like them we also mask the Sagittarius dwarf galaxy which reduces 
the original sample size by 40$\%$ to $\sim 3500$. 

D11 measure the anisotropy of the halo in radial and metallicity bins.
In the inner halo ($10 < r/\kpc< 25$) both metal-rich ([Fe/H]$>-2$) 
and metal poor ([Fe/H]$<-2$) stars are found to be tangential with
$\beta$ $\sim-1.2$ and $\sim-0.2$ respectively. In the corresponding
metallicity bins, the outer halo ($25 < r/\kpc < 50$) is found 
to be radial with $\beta$ of $\sim0.4$ and $\sim0.5$ respectively.
Since they do not give the estimates of $\beta$ in combined metallicity bins
for the inner and outer halo, we estimate them here using the same 
methodology as adopted by them. 
For the inner halo we find $\beta=-0.62$ (tangential) and for the outer halo 
we find $\beta=0.41$ (radial). 
These estimates are consistent with D11 results, 
if we combine their low and high metallicity $\beta$ values in each radial bin.
 
The inner halo within the solar vicinity has been found to be radial in 
studies of halo subdwarfs by \citet{2009MNRAS.399.1223S} and in studies of 
$10^{5}$ main sequence stars by \citet{2010ApJ...716....1B}
(see also Figure \ref{fig:aniso}{d}). 
Hence, a tangential inner-halo as predicted by  D11 is surprising. 
The first thing to check is if the D11 result is due to some of the 
assumptions made by them. For example, in D11 
the logarithmic density slope was assumed to be constant and equal to $-3.5$. 
Later on \cite{2011MNRAS.416.2903D} conducted a detailed 
analysis of the  BHB stars to estimate their density profile 
and found that the profile is of the form of a 
broken power law, the inner-halo ($<27$ kpc) having 
a profile index of $-2.3$ and the outer halo having a profile index of $-4.6$.
This is in good agreement with the findings of \citet{2009MNRAS.398.1757W} 
that the halo within 25 kpc has the profile index $-2.4$ and beyond 
which it is steeper with an index $-4.5$. 

If we update the density profile index in the D11 case with the above values 
then we expect the inner halo which is already tangential
to become even more tangential and the outer halo which is
already radial to become even more radial. 
This is because $\beta$ has a dependence on the adopted value of $\alpha$ 
as shown by D11 (Figure 5). 
In general $\beta$ increases with an increase in the  adopted value of $\alpha$.
Another effect that can potentially bias the results is the fact that the potential parameters 
($\Phi_{0}, \gamma$) have been kept fixed in the 
D11 analysis, i.e., the degeneracy between   $\beta$ 
and potential has not been broken. 
After analyzing data in finer bins and breaking the degeneracy among the 
model parameters we do see that in Figure \ref{fig:beta_2models} 
the inner halo is radial (from \S \ref{sec:alternative_model}) as was also found with the GVE model. 
Hence the assumption of a fixed potential can bias the estimation of $\beta$. 
However, a much more apparent reason for the discrepancy is as follows.
Clearly, from Figure \ref{fig:aniso} the radial bin from 10-25 kpc will encompass the stars within 
13-17 kpc which are predominantly tangential.
Since the probability density of stars in radius also peaks at 
around 16 kpc (see Figure \ref{fig:rgc}) we anticipate the 
overall $\beta$ to be tangential.  
To conclude,  the tangential behavior of the 
inner-halo seen by D11 is most likely due to the 
large radial bin size adopted by them.

\subsection{Anisotropy at $16 < r/\kpc< 48$ seen by D12}
\label{subsec:r1648}

\begin{figure}
    \includegraphics[width=0.4\textwidth]{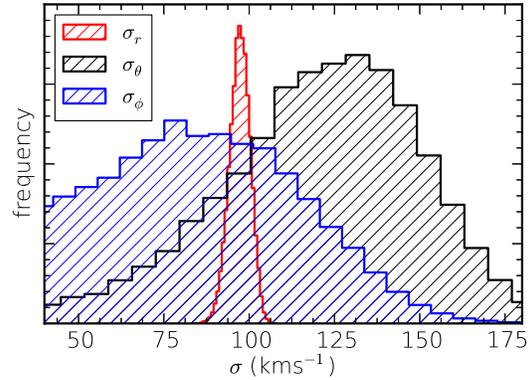}
  \caption{Posterior distribution of velocity dispersions for D12 
           data set within $16 < r/\kpc< 48$  
           using a GVE model. The value of v$_{\rm LSR}$ here 
           is taken to be 240 kms$^{-1}$ to keep it same as in D12. }
\label{fig:vel_dispersion_r1648}
\end{figure}

D12 re-calibrate the distances of BHB stars using the 
color-magnitude relation given by \cite{2011MNRAS.416.2903D} and then select stars within 
$16 \lesssim r/\kpc \lesssim 48$ ( 1933 stars) 
from \citet{2011ApJ...738...79X} BHB samples. 
They fit the D11 model  to study the nature of $\beta$ in the outer halo. 
Unlike D11, as mentioned previously here they
break the degeneracy present in their model and consider the $\beta$, $\Phi_{o}$ 
and $\gamma$ as free parameters. Thus while they fit the model they 
simultaneously estimates all three parameters. 
They find $\beta=0.4^{+0.2}_{-0.2}$ for $\alpha=4.6$. 
Using a model allowing for oblateness ($q=0.59$) they find $\beta=0.5^{+0.1}_{-0.2}$.  
If we apply the GVE distribution function to stars in the range 
$16<r/\kpc<48$ we find $\beta=-0.14^{+0.52}_{-0.66}$ ($\rsigma=97.3^{+2.9}_{-3.0}$ kms$^{-1}$, 
$\tsigma=122.7^{+26.4}_{-33.0}$ kms$^{-1}$ and $\psigma=78.5^{+34.3}_{-40.7}$ kms$^{-1}$). 
This more or less looks like the mean value of $\beta$ in this range 
(see Figure \ref{fig:beta_2models}) provided we take into account 
the fact that the number density of stars peak at around $r=16$ kpc.  
Although the D12 value is still within our 1 $\sigma$ region, our  
predicted value is lower than D12. 
It can be seen from Figure \ref{fig:beta_2models} that 
$\beta$ is not constant in the range $16<r/\kpc<23$.  
It increases from being tangential to isotropic. 
Using the D11 DF also gives similar result (\S \ref{sec:alternative_model}). 
Beyond this range we cannot directly measure $\beta$, but by deriving a 
best fit circular velocity profile and making use of $\rsigma$ 
profile which is available till $r=56$ kpc we can predict $\beta$, 
and this is returned to in section \ref{sec:circ_velocity}. 
However, an assumption about $\alpha$ also has to be made.
Beyond, $r>27$ kpc the density slope has been shown to change 
from $-2.4$ to $-4.5$. Adopting a steeper density slope increases the value 
of $\beta$. For $\alpha=4.5$ we find $\beta \sim 0.5$ for $r>27$ kpc; 
this is more in agreement with D12. To conclude, 
the D12 value of $\beta=0.4$ for $16<r/\kpc<48$, 
although derived for a sample which is dominated by stars within 
$r<27$ kpc, is not appropriate for the range $18<r/\kpc<23$, instead 
it is correct for the range $23<r/\kpc<48$.

\begin{figure}
    \includegraphics[width=0.45\textwidth]{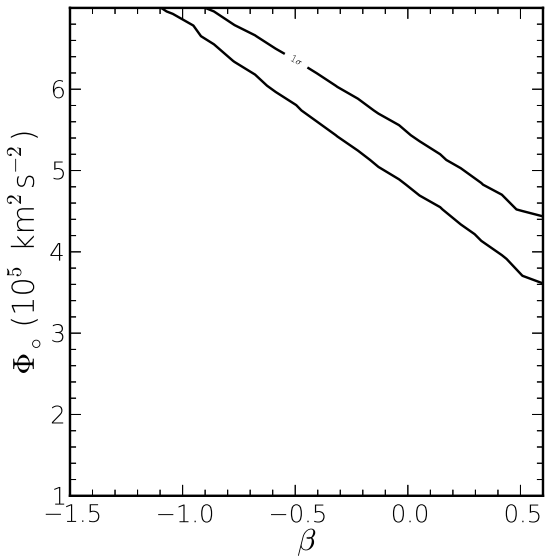}
  \caption{Maximum likelihood analysis of the anisotropy 
           parameter, $\beta$, for stars in the radial 
           bin $35 < r/\kpc <  84$. Solid black lines are the 1$\sigma$
           confidence region. Density and potential slopes are taken
           to be 4.5 and 0.35.}
\label{fig:r35_84_D11}
\end{figure}

Finally, we check how best we can measure $\beta$ in the outer 
most parts, $35 < r/\kpc < 84$, using the D11 DF and the data in hand. 
This region consists of $762$ stars, and we assume  $\alpha=4.5$ and $\gamma=0.35$ and repeat 
the analysis as in D12. The likelihood distributions of model 
parameters are shown in Figure \ref{fig:r35_84_D11}. 
The mass-anisotropy degeneracy is clearly visible here, suggesting 
that it is very difficult to directly measure $\beta$ 
unless an explicit or implicit assumption about the potential 
is made.

\section{Circular Velocity curve of the Galaxy}
\label{sec:circ_velocity}
\begin{figure*}
    \includegraphics[width=0.95\textwidth]{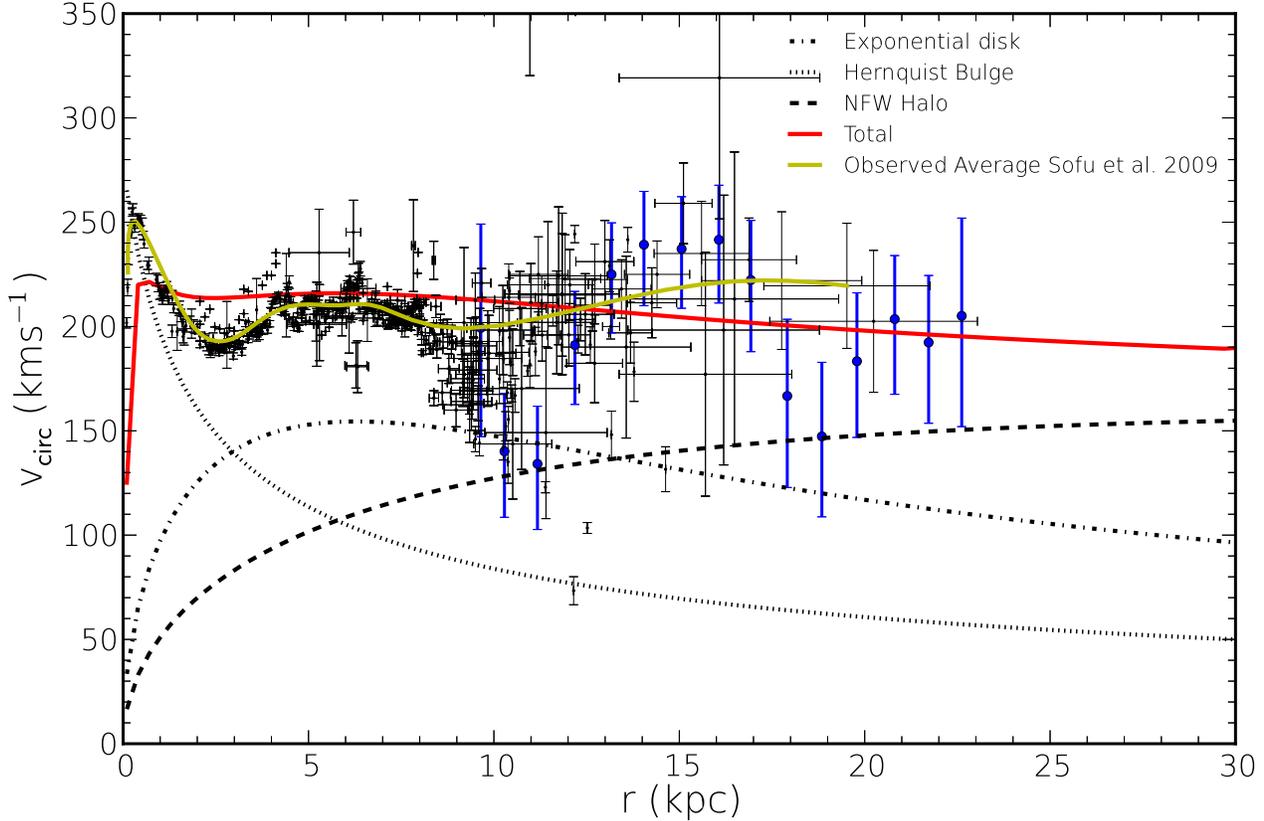} 
  \caption{Circular velocity curve of the Galaxy and their individual
           components along a galactocentric distance ($r$). 
           The blue marker represents the value of $\vcirc$
           obtained in the CME bins in $r$. Red solid line is our fit 
           of the total potential. Black dotted and dotted-dashed lines are 
           the fixed disk and the bulge circular velocity profile for set of adopted 
           values of masses and scale radii. Dashed line is the fitted NFW profile.
           Black dots with error bars are the collated $\vcirc$ values
           given by \citet{2009PASJ...61..227S} whereas yellow solid line is the average of the
           given observed values.}
\label{fig:vcirc}
\end{figure*}

Here we use the measured values of $\rsigma(r)$ and $\beta(r)$ from our analysis 
given in Figure \ref{fig:aniso} to determine the circular velocity curve of 
the Galaxy ($\vcirc$) through the Jeans equation (Equation \ref{eqn:Jeanseq}).  
Besides anisotropy and radial velocity dispersion information
we also need to adopt some density profile for the tracer population
but not of the spectroscopic sample.  
We adopt a value of  $\alpha$ = 2.4 as suggested by recent works of 
\cite{2009MNRAS.398.1757W} and \cite{2011MNRAS.416.2903D}, for 
the range of distance explored here ($r<25$ kpc).  
The blue dots with error bars in Figure \ref{fig:vcirc} are our estimates
of $\vcirc$ using CME with $n_{\rm bin} = 1200$.
The uncertainties on $\vcirc$ were computed using a Monte Carlo 
based scheme from uncertainties in $\rsigma$ and $\beta$. 
The leftmost and rightmost points have comparatively larger error bars 
as compared to intermediate points. For the leftmost point the large error 
bar is mainly due to large error in the value of $\rsigma$. On the
other hand, for the right-most point the large error bar is mainly  
due to large error in the values of $\tsigma$ and $\psigma$.

In Figure \ref{fig:vcirc} it can be seen  that the circular velocity 
profile derived from our analysis (blue points) display prominent features. 
We now check if such features are also observed in other studies 
using tracers other than BHB stars. 
For this we over-plot $\vcirc$ compiled by 
\cite{2009PASJ...61..227S} as black dots, obtained from several 
references \citep{1978A&A....63....7B, 1982ApJS...49..183B,
1985ApJ...295..422C, 1989ApJ...342..272F, 1997PASJ...49..453H, 1997PASJ...49..539H,
2007PASJ...59..889H, 2007A&A...473..143D}. 
Further details about the source of each individual point  
can be found in \citet{2009PASJ...61..227S} and references therein.
Note, $\vcirc$ values of \cite{2009PASJ...61..227S}
are computed for (R$_{\sun}$, v$_{\rm LSR}$) = (8.0 kpc, 200.0
kms$^{-1}$). 
Correcting the $\vcirc$ for our adopted values of (R$_{\sun}$, v$_{\rm LSR}$)
= (8.5 kpc, 220.0 kms$^{-1}$) is beyond the scope of this work and
hence we simply over-plot these published values in Figure \ref{fig:vcirc}. 
It can be seen there is a prominent dip at 9 kpc in the \cite{2009PASJ...61..227S} 
compiled $\vcirc$ profile. They explain this dip by 
introducing massless rings on top of a disk with exponential surface density.
We also find a similar dip in our $\vcirc$ profile 
but at around 10-12 kpc. The slight shift in the position of dip could 
be due to large width of our bins and also due to the fact that unlike
\cite{2009PASJ...61..227S}, who measure $\vcirc$ in the mid-plane 
of the Galaxy, we measure $\vcirc$ over a spherical shell
that intersects with the SDSS footprint. 

We now estimate the mass of the dark matter halo of the Milky Way, 
assuming a three component model of the Galaxy consisting of the
bulge, the disk, and the halo. The bulge is modeled as a Hernquist
sphere and the disk is assumed to follow an exponential profile \citep{2008ApJ...684.1143X}. 
The parameters for the bulge and disk are taken from \cite{2009PASJ...61..227S}
and are kept fixed. Although \cite{2009PASJ...61..227S} use massless rings, we here 
have ignored them since our main aim is to fit the dark matter 
halo. We model the dark matter halo using the NFW 
\citep{1996ApJ...462..563N} density profile. Here we consider both the
halo and the bulge to be spherically symmetric. 
The non-axisymmetric effect due to a bar shaped bulge is neglected here.
Potentials for bulge, halo, and exponential disk can be expressed as,
\begin{eqnarray}
\label{eqn:NFWpot}
\Phi_{\rm disk} (r)& = &-\frac{GM_{\rm disk}(1-e^{-r/b})}{r} \label{eqn:diskpot},\\
\Phi_{\rm bulge} (r)& = &-\frac{GM_{\rm bulge}}{r+a} \label{eqn:bulgepot},{\rm and}\\
\Phi_{\rm NFW} (r)  & = &-\frac{GM_{\rm vir}\ln(1+r c/R_{\rm vir})}{g(c)r} 
\end{eqnarray}
where  $M_{\rm disk}$ = 6.5$\times$10$^{10}$ $\msun$, $b$ = 3.5 kpc, 
$M_{\rm bulge}$ = 1.80$\times$10$^{10}$ $\msun$, and $a$ = 0.5 kpc
\citep{2009PASJ...61..227S}.

Note, the disk potential as given by Equation (\ref{eqn:diskpot}) 
is spherically symmetric. 
It means the disk is considered to be a
spherical body with exponential surface density fall-off. 
To get an idea on the error that is incurred due to the assumption of
the disk as a spherical body with the mass same as the flattened disk 
we refer reader to \cite{2008gady.book.....B}(Figure 2.17). 
Roughly the maximum error in $\vcirc$ is $13\%$, which is at a distance about 
twice of the disk scale length. 
However, at the larger distances along the mid-plane the discrepancy is smaller. 
In the general case of triplanar symmetry (elliptic disk), in reality, the disk
potential has to be the function of both polar and azimuthal coordinates
and in the special case of axial symmetry (circular disk) it has to be
the function of sole polar coordinates, in addition to the radial coordinates ($r$).  
We here use the spherically symmetric form for two reasons.
Firstly, we make use of the the spherical form of the Jeans equation 
given by Equation (\ref{eqn:Jeanseq}) which demands a spherical 
potential. Secondly, it is to ease the comparison with earlier studies 
, e.g., \citet{2008ApJ...684.1143X}, that adopt a similar definition.
However, later on we consider a 3D disk potential 
and discuss its consequences on the estimation of mass.
The function $g$ in NFW potential is given by \[g(c) = \ln(1+c) - \frac{c}{1+c}\] and 
\[R_{\rm vir} =\left(\frac{2 M_{\rm vir}G}{H_0^2 \Omega_m \Delta_{\rm th}}\right)^{1/3}.\]

The total potential $\Phi(r)$ of the Galaxy is then simply
\begin{equation}
\label{eqn:totalpot}
\Phi(r) = \Phi_{\rm bulge}(r) +  \Phi_{\rm disk}(r) +  \Phi_{\rm NFW}(r)
\end{equation}

We adopt the value of Hubble constant, H$_{0}$ = 70.4 kms$^{-1}$Mpc$^{-1}$, 
$\Omega_{\rm m}=0.3$ \citep{2011ApJS..192...18K}, and
$\Delta_{\rm  th}=340$ \citep{1998ApJ...495...80B}.

A NFW halo has two free parameters the mass $M_{\rm vir}$ and the
concentration $c$. Since we do not have enough data points spanning 
a wide range in radius, we avoid fitting both the parameters
simultaneously. Instead we use the concentration mass relation, 
\begin{equation}
c = 327.3 M_{\rm vir}^{-0.12}, \ \  10^{11} \leqslant M_{\rm vir}/M_{\sun} \leqslant 10^{13},
\end{equation}
as has been reported in N body simulations of dark matter halos \citep{2007MNRAS.378...55M}. 

Finally, we can derive the resultant circular velocity ($\vcirc$) from the total potential
(Equation \ref{eqn:totalpot}) by computing $(r d\Phi/dr)^{1/2}$. 
We fit the obtained theoretical rotation velocity curve to 
our observed values of $\vcirc$ shown by the blue dots in Figure \ref{fig:vcirc}, 
and the red line is our best fit circular
velocity curve. The $\vcirc$ profiles for the
different components are also shown separately. 
The dashed black line is the corresponding best fit NFW halo
profile. 
The best fit value for the fit parameter, $M_{\rm vir}$,
for our three component baryon and dark matter mass distribution 
is 0.9$^{+0.4}_{-0.3}$ $\times$ 10$^{12}$ $\msun$. The corresponding values of  
$R_{\rm vir}$ and $c$ derived from the best fit value of $M_{\rm vir}$ 
are 249$^{+34}_{-31}$ kpc and 12.0$^{+0.6}_{-0.5}$ respectively.
We estimate the mass of the Galaxy within $r \lesssim 25$ kpc
to be 2.1 $\times$ 10$^{11}$ $\msun$. Assuming a functional form 
for $P(v_{los}/\vcirc)$ obtained from simulations \cite{2008ApJ...684.1143X} 
derive the $\vcirc$ from the  $v_{los}$ of BHB stars. The derived $\vcirc$ 
is then  used to estimate the virial mass of the dark matter halo. 
They find $M_{\rm vir} = 0.91^{+0.27}_{-0.18} \times 10^{12}$ $\msun $ 
which is in good agreement with our result, note  
uncertainties are however, slightly larger in our estimates.
Since unlike them we do not make any assumption about the functional 
form of $P(v_{los}/\vcirc)$.   

Here we study the effect of chosen disk models for which
we consider more realistic three-dimensional potential 
for the disk by \cite{1975PASJ...27..533M}, 
which holds for the special case of a circular disk.
The expression for this potential is given by
\begin{equation}
\label{eqn:miyamotondisk}
\Phi_{\rm disk} (R,z) = -\frac{G M_{\rm disk}} {\sqrt{R^2+(a+\sqrt{z^2+b^2})^2}}.
\end{equation}
Here again, the disk parameters are obtained 
from the best fit values along the galactic mid-plane $(z=0)$ 
which reproduces $\vcirc(R)$ profile for \cite{2009PASJ...61..227S}
razor-thin exponential disk (\citealt{1970ApJ...160..811F})
i.e. $b = 0.0\:\kpc$ \footnote{The disk model with $b=0$ is also otherwise known as 
Toomre's model or Kuzmin disk \citep{1963ApJ...138..385T, 2008gady.book.....B}}. 
Hence the best fit value for the disk parameter $a = 2.5$ kpc whereas 
mass $M_{\rm disk} = 6.5 \times 10^{10}$ $\msun$ is taken same as
in \cite{2009PASJ...61..227S}.
Since we assume the disk potential to be three-dimensional here,
for the purpose of computing the total $\vcirc$ we consider 
the component of force along the radial direction ($r$) only. 
The bulge and the halo models are kept same as earlier.
The best fit values for the NFW halo parameters with the updated
disk model is found to be $M_{\rm vir} = 1.2^{+0.5}_{-0.4} \times 10^{12}$ $\msun$ with
$R_{\rm vir} = 274^{+35}_{-30} \: \kpc$. 
Instead if we consider the total magnitude of the
force in order to compute the $\vcirc$ we estimate 
$M_{\rm vir} (R_{\rm vir} = 269^{+34}_{-32}\:\kpc) = 1.1^{+0.5}_{-0.4}
\times 10^{12}$ $\msun$. 
Historically, large values of $a$ have been assumed 
while modeling the disk (\cite{1995ApJ...453..673W, 2009IAUS..254..241B}). We find that this leads to a much 
more massive dark matter halo-- for $a = 4.5 \: \kpc$  we find  
$M_{\rm vir} (R_{\rm vir} = 299^{+36}_{-33}\ \kpc) = 1.6^{+0.6}_{-0.5} \times 10^{12}$ $\msun$ 
and for $a=6.5\:\kpc$ we find 
$M_{\rm vir} (R_{\rm vir} = 321^{+35}_{-34}\:\kpc) = 1.9^{+0.7}_{-0.6} \times 10^{12}$ $\msun$.

There are few things which have insignificant or unexplored effects on
our mass estimation e.g. we do not take into account the mass of the 
super-massive black hole, which is $\sim 4\times10^6$ $\msun$ 
\citep{2002Natur.419..694S,2005ApJ...628..246E,2009ApJ...692.1075G},
and is approximately 1/1000 of the mass of the bulge.  
Moreover, its effect is like that of a point mass and can be 
easily absorbed into the bulge mass.
Another effect that is not considered 
is the tidal effect on the primary object that has been qualitatively studied to 
find the mass ratio between the Galaxy and M31 \citep{2009MNRAS.397.1990B}
and depends strongly on different impact parameters \citep{1973A&A....22...41E}.
We find that R$_{\rm vir}$ of the Galaxy is $\sim250$ kpc
and for M31 it is $\sim 260$ kpc 
\citep{2008MNRAS.389.1911S, 2007ApJ...670L...9M}.
Given that the distance between these two galaxies is 
$\sim780$ kpc \citep{2005ApJ...635L..37R, 2005MNRAS.356..979M, 2006Ap.....49....3K}, 
which is more than the double of the sum of their virial radii, we believe the tidal effect
of M31 on overall mass estimate of the Galaxy, if any, should be negligible.
The tidal effects of LMC and SMC on the Galaxy have not been explored in this paper.

\begin{figure}
    \includegraphics[width=0.475\textwidth]{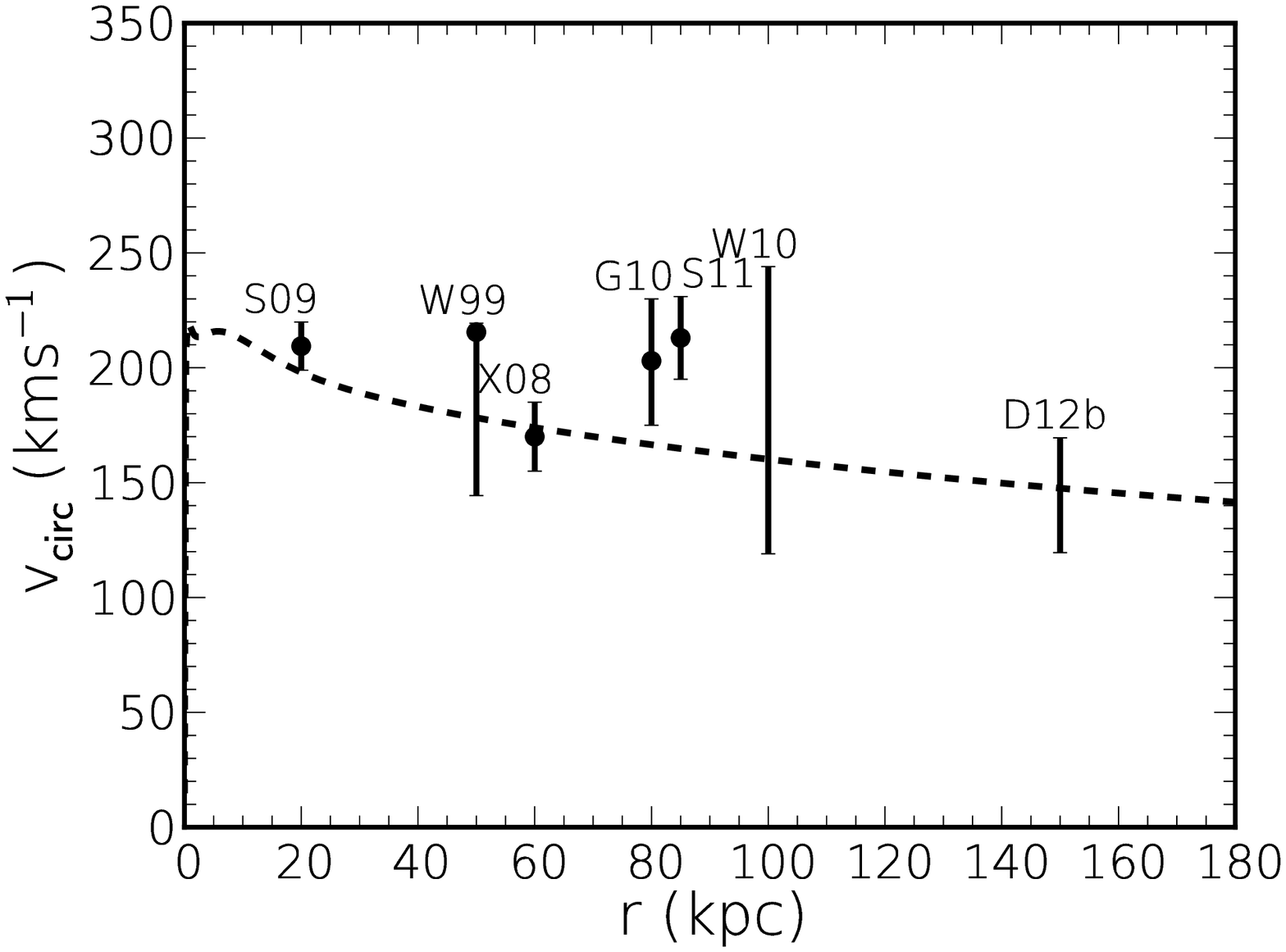} 
  \caption{Dashed black line is our best fit model of $\vcirc$
           given by red line in Figure \ref{fig:vcirc}.
           Black dots are the values from the literatures 
           \citep{1999MNRAS.310..645W, 2008ApJ...684.1143X, 2009PASJ...61..227S,
           2010MNRAS.406..264W, 2010ApJ...720L.108G, 2011A&A...531A..82S, 2012MNRAS.425.2840D} labeled
           as W99, X08, S09, W10, G10, S11, and D12b respectively. To make the plot less obscure 
           we do not include similar findings from the literature. For details about 
           the similar results refer to the text.}
\label{fig:vcirc_all}
\end{figure}

Note, our mass modeling of the Milky Way, does not make any
assumption about the value of $\beta$, instead we use the value of $\beta$  
directly computed from the data. The only assumption that we make is 
that the density of the dark matter 
halo follows an NFW profile. As long as that 
assumption holds our estimates for $\vcirc$ and mass of the Milky 
Way should also be valid in the outer parts $r>25$ kpc where   
we cannot directly measure $\beta$. 
If one wants to directly measure $\vcirc$ in the outer parts  
using only line of sight velocities then one has to make 
an assumption about the underlying $\beta$.
Several attempts have been 
made in this regard, with each of them making different assumptions 
about $\beta$ and hence introducing a bias in the estimated mass. 
Below we compare these with our prediction for the mass
of the Galaxy.
Dashed line in Figure \ref{fig:vcirc_all} is our best fit model
of $\vcirc$. 
Since literature sources mostly report mass within a 
certain radius, to facilitate comparison we convert it to  $\vcirc$  using the relation 
\[M(<r) = \vcirc^{2}r/{\rm G}.\]
In Figure \ref{fig:vcirc_all} the plotted $v_{\rm circ}$ from different
sources span a wide range in radii 
and were computed using different types of tracer populations. 
By fitting a model to the kinematics of the 
satellite galaxies and the globular clusters, \cite{1999MNRAS.310..645W} measure the mass 
to be $M(50 \:\kpc) \sim 5.4^{+0.2}_{-3.6} \times 10^{11}$ $\msun$. This agrees 
with estimates by \cite{1996ApJ...457..228K} ($M(50 \ \kpc) = (4.9 \pm 1.1) \times 10^{11}$ $\msun$) 
and \cite{2003A&A...397..899S} ($M(50 \ \kpc) = 1.8-2.5 \times 10^{11}$ $\msun$).
\cite{2010MNRAS.406..264W} apply the tracer mass estimator formalism 
to 26 satellite galaxies and find that $M(300 \ \kpc) = (0.9 \pm 0.3) \times 10^{12}$ $\msun$.
Their mass estimate is however prone to the systematics introduced by assumed $\beta$,
as duly mentioned by them. It can be seen in the figure that at $r$ = 100 kpc, 
depending upon the chosen anisotropy, their mass could vary anywhere between 
$0.3 \times 10^{12}$ $\msun$ and $1.4 \times 10^{12}$ $\msun$.
Studying the hyper-velocity stars within 80 kpc, and assuming 
$\beta=0.4$, \cite{2010ApJ...720L.108G} estimate $M = 6.9 _{+3.0}^{-1.2} \times 10^{11}$ $\msun$, 
which  is slightly higher than our estimate. 
Using BHB stars and \cite{2010MNRAS.406..264W} tracer formalism 
\cite{2011A&A...531A..82S} estimate the mass at $r = 85\ \kpc$ 
to be $8.83\pm{0.73}\times 10^{11}$ $\msun$ . This is slightly 
higher than our estimate, probably because they assume $\beta=0$.  
With the mixed sample of tracers
(BHB and CN stars) populating the outer-most halo ($r\sim50-150\:\kpc$),
\cite{2012MNRAS.425.2840D} estimated mass of the Galaxy to be $M(150 \: \kpc)=(5-10) \times 10^{11}$ $\msun$. 
The variation is mainly due to uncertainty on the adopted 
potential and density slopes, and anisotropy.
Their range of mass at the outer-most halo falls within our estimation.

\section{Kinematics of the simulated stellar halo}
\label{sec:simulated_halo}
\begin{figure}
    \includegraphics[width=0.475\textwidth]{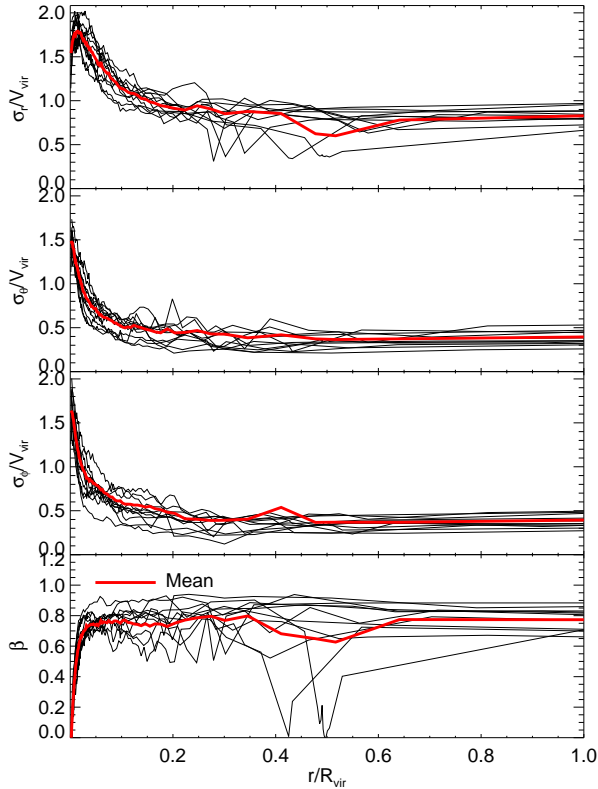} 
  \caption{Velocity dispersions and anisotropy profiles of BHB stars
    in simulated stellar halo:
           From top to bottom are $\rsigma$, $\tsigma$, $\psigma$, and $\beta$
           profiles of the 11 instances of simulated halo taken from 
           \cite{2005ApJ...635..931B}. The thick red lines are the mean 
           profiles of the 11 halos.}
\label{fig:kinematics_simulation}
\end{figure}

\begin{figure}
    \includegraphics[width=0.475\textwidth]{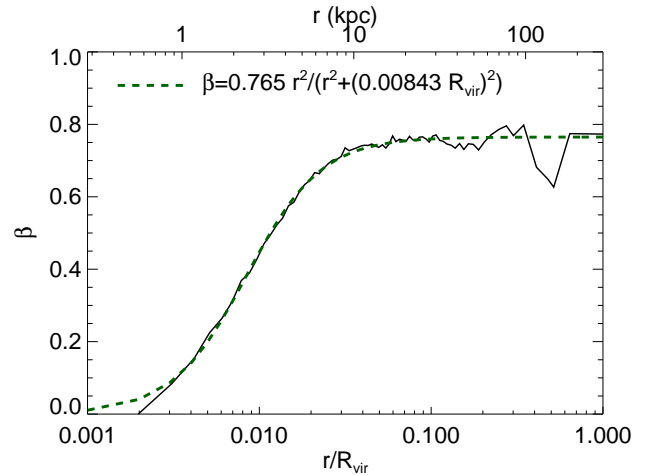} 
  \caption{The mean $\beta$ profile of BHB stars in 11 simulated $\Lambda$CDM
    stellar halos of \cite{2005ApJ...635..931B}. Shown alongside as
    dashed line is  the best fit analytic function of form given by 
    \cite{1991MNRAS.253..414C}.}
\label{fig:beta_lcdm}
\end{figure}

\begin{figure}
    \includegraphics[width=0.475\textwidth]{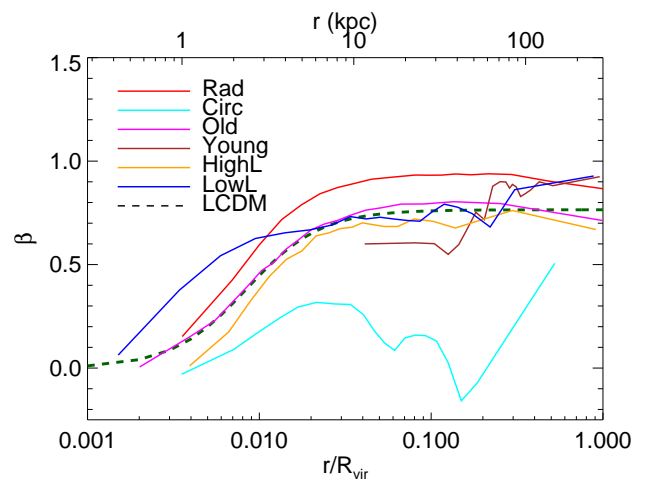} 
  \caption{The $\beta$ profile of simulated stellar halos having 
non-standard accretion history.}
\label{fig:beta_new}
\end{figure}

We now study the kinematic properties of simulated stellar halos 
in which the halos are formed purely by accretion of satellite
galaxies. For this we use the simulations of
\cite{2005ApJ...635..931B}. In order to construct a synthetic 
sample of BHB stars from these simulations we use the  
code GALAXIA \citep{2011ApJ...730....3S}.
Figure \ref{fig:kinematics_simulation} shows the velocity dispersion 
and anisotropy profiles of 11 different $\Lambda$CDM halos as a function 
of galactocentric radius $r$. 
The mean of all the halos is also shown alongside as thick red line. 
In general the velocity dispersions fall off with radius. At small $r$ the 
fall is rapid but at large $r$ it is much slower. Asymptotically 
the ratio $\rsigma/V_{\rm vir}$ approaches a value of around 0.8.
The $\beta$ rises rapidly from a value of zero in the center to about 0.8 
at $r\sim 10 \:\kpc$, and thereafter shows very little change.
These results are in good agreement with results of 
\cite{2006MNRAS.365..747A, 2007MNRAS.379.1464S}, who study the stellar halo formed 
in cosmological hydro-dynamical simulations including star formation 
and feedback. 

Firstly notice that $\Lambda$CDM halos are rarely tangential 
for any given radius. For most of the range of $r$, $\beta$ is 
in general greater than 0.5. We fit an analytic function of 
\cite{1991MNRAS.253..414C} form given by 
\begin{equation}
\beta(r)=\beta_0 \frac{r^2}{r^2+r_0^2}
\label{eqn:cford_profile}
\end{equation}
to the mean $\beta$ profile of the 11 $\Lambda$CDM halos. The best fit
values of the free parameters were found to be 
$\beta_0=0.765$ and $r_0=0.00843 R_{\rm vir}$. Figure
\ref{fig:beta_lcdm} shows that the fit is quite good for a wide range
of $r$. The slight mismatch at $r<1$ kpc could be due to issues related 
to force resolution. In the outer parts most of the mass is in bound
structures and hence is not smoothly distributed. This is probably 
responsible for the non-smooth behavior in $\beta$ in the outer parts.

Figure \ref{fig:beta_new} presents the beta profiles for halos
having non-$\Lambda$CDM accretion history, i.e., halos having accretion
history significantly different from that predicted by the $\Lambda$CDM model 
of galaxy formation. 
Six halos that we consider are with accretion events being 
dominated by 1) radial orbits 2) circular orbits 3) old events 4) recent events 
5) higher luminosity and 6) low luminosity and these are from 
simulations of \cite{2008ApJ...689..936J}. 
Signatures of different accretion events can be seen in the 
$\beta$ profiles. 
The most significant difference is between the radial and the circular halo, 
which is expected since the orbital properties of the 
satellites were different to begin with. 
It is interesting to note that the circular halo is the only one  
among all the simulated halos that can have $\beta<0$. The old halo 
almost perfectly follows the mean profile that we had derived for 
the 11 $\Lambda$CDM halos and has the smoothest profile. This is due to the 
fact that the stars in this halo are completely phase mixed and have 
no structures of any kind. 
The young halo has very few stars in inner regions 
and shows non smooth behavior due to presence of significant
amount of structures.  
The high luminosity halo is also very similar to 
old halo. However, the low-luminosity halo has higher $\beta$ for $r<0.01 R_{\rm vir}$. 
This is most likely due to circularization of orbits when
acted upon by dynamical friction. 
Orbit circularization has also been reported by \cite{2007MNRAS.379.1464S} in their 
simulations. 
Note, the effect of dynamical friction is strongest for 
high luminosity events and weakest for low luminosity events. 
Moreover, satellites when acted upon by dynamical friction 
loose energy and move towards the inner regions of the halo. 
This partly explains as to why the high luminosity halo 
has low $\beta$ in the inner regions as compared to the low luminosity
halo.

\begin{figure*}
    \includegraphics[width=0.850\textwidth]{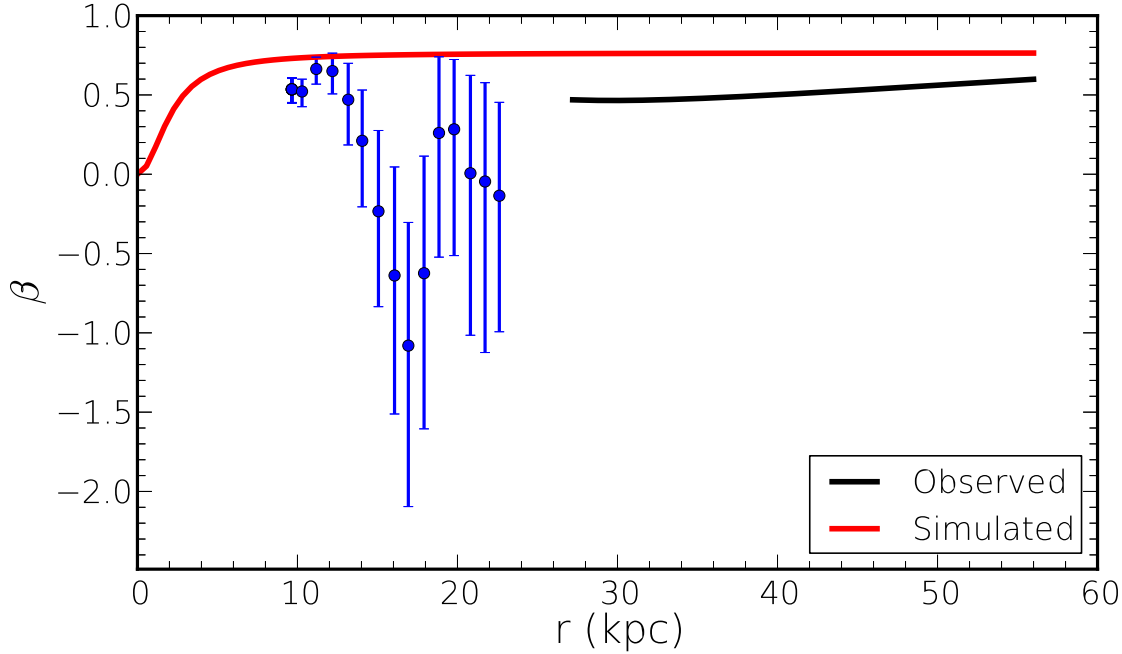} 
  \caption{Black solid line is the observed $\beta$ profile in the outer halo
           (assumed $\alpha = 4.5$) whereas the red solid line is the $\beta$ profile 
           of the simulated halo. The blue dots with error bars are the observed values of $\beta$ from 
           section \ref{sec:dispersion_profile}. The sudden jump in $\beta$ profile 
           passing from $r = 23$ kpc to $r = 27$ kpc could be 
           a spurious effect due to our assumption of the broken power law 
           with break at $r=27$ kpc.}
\label{fig:beta_prediction}
\end{figure*}

In \S \ref{subsec:tangentialdis} we measured $\beta$ until $r=23$ kpc.  
Beyond this we can measure $\rsigma$ out to $r=56$ kpc, but not 
$\psigma$ and $\tsigma$. By using the circular velocity 
curve that we derived in \S \ref{sec:circ_velocity} we can 
predict $\beta$ beyond $r>23$ kpc making use of the 
Jeans equation (Equation \ref{eqn:Jeanseq}). 
To proceed we need to make an assumption about the density slope ($\alpha$); 
beyond $r>27$ kpc it has been shown that $\alpha$ is around 4.5.
Assuming this, the predicted $\beta$ is plotted in Figure
\ref{fig:beta_prediction}.  
It can be seen that there is a slight 
jump in the value of $\beta$ passing from $r = 23$ kpc to
$r = 27$ kpc and beyond this the value of $\beta$ is around 0.4.
The sudden jump in $\beta$ occurs via 
the Jeans equation (Equation \ref{eqn:Jeanseq}), due to the discontinuity 
in $\alpha$ ($= - {\rm d} \ln \rho/ {\rm d} \ln {\rm r}$).
Note, an assumption of steeper density slope would increase the $\beta$ and vice versa.
The red line in figure is the anisotropy profile 
fitted to the simulated $\Lambda$CDM halos given by Equation 
(\ref{eqn:cford_profile}) from \S \ref{sec:simulated_halo}. 
It can be seen that accretion based models cannot explain 
the dip that is in observations, specially the profile in the 
region $12<r/\kpc<23$. However, outside this region the simulations 
are roughly in agreement with observations.
For $r<12$ kpc the observations match the value of around $\beta=0.5$ 
seen in simulations. For $r>23$ kpc 
the overall value of $\beta$ in observations 
is slightly low but the profile is flat as in simulations.
The outer halo at $r=56$ kpc is radial with $\beta=0.55$.

\section{CONCLUSION and DISCUSSION}\label{sec:conclusion}
We study the kinematics of $\sim4500$ BHB stars to 
obtain the velocity dispersion profiles along three orthogonal axes
in spherical polar coordinates using the gaussian velocity ellipsoidal 
(GVE) DF. 
GVE as an estimator of the velocity dispersion 
has the advantage that no assumptions about potential or density are needed {\it a priori}. 
From the estimated velocity dispersion profiles using maximum likelihood 
analysis, we also derive the anisotropy profile of the Galactic halo and 
compare it to the simulated $\Lambda$CDM halos. 
Finally, using radial velocity dispersion profile, 
anisotropy profile and density power law we constrain the mass of the 
Milky Way Galaxy using the Jeans equation. 

We measure the $\rsigma$ profile of the halo out to $r\sim60$ kpc. 
At large distance ($d\gg R_{\sun}$), $\rsigma$ can be approximated by $\sigmalos$.
Thus in outskirt, $\sigmalos$ profile given by \cite{2008ApJ...684.1143X}
converges to our $\rsigma$ profile. At $r\sim60$ kpc 
$\rsigma$ attains $\backsimeq 100$ kms$^{-1}$. 
However, in the inner halo the approximation ($\rsigma\approx\sigmalos$) is invalid and 
we find that the deviation of $\sigmalos$ from $\rsigma$ is as high as $\sim40$ kms$^{-1}$. 
We obtain a $\rsigma(r)$ profile with plateau in the inner halo,
a sudden fall at $r\sim15$ kpc and a gradual decline outwards.
Qualitatively, similar profile is also found by \citet{1997ApJ...481..775S}.
However, our $\rsigma$ profile sharply falls at $r\sim15$ kpc
whereas they find a gentle transition. 
The is probably due to the fact that they make an assumption that 
$\vcirc(r)$ is constant which we find is not completely true.

Next we estimate the tangential velocity dispersions, $\tsigma$ and $\psigma$.
Using these estimates we are able to measure the $\beta(r)$ till $r = 25$ kpc.
Astonishingly, we discover a dip in the $\beta$ profile at $r = 17$ kpc, 
where $\beta\backsimeq-1.2$. We find that the inner halo ($r<12$ kpc) 
is radial with $\beta \backsimeq 0.5$. This result of radially biased inner halo
concurs with the recent results by \citealt{2009MNRAS.399.1223S, 2010ApJ...716....1B} using the proper motions.
Beyond the switch over point in the range 
$18 \lesssim r/\kpc \lesssim 25$ the anisotropy rises slightly
and becomes isotropic to mild radial. 
We also verify the result using an alternative DF, namely the D11 DF. 
A small systematic in the $\beta(r)$ profiles is seen from 
these two models which is mainly due to the assumption about density slope 
($\alpha$) that needs to be made {\it a priori} for D11 DF.
We check for the contribution of the halo substructures,
namely Virgo Over-density and Sagittarius stellar stream, and find that they have little 
effect in the anisotropy profile.
The effects of v$_{\rm LSR}$ and R$_{\sun}$ upon our velocity dispersions and anisotropy 
estimates are also found to be negligible.

D11 study the BHB stars in the radial bin ($10<r/\kpc<25$) and find the halo 
to be tangential. After re-analyzing the D11 sample in this bin, 
we find that this is mainly because of the choice of their bin size that encompasses the transition 
region ($13<r/\kpc<17$) where we detected a dip in the $\beta(r)$. 
Possibly it could be also because of the degeneracy between potential and anisotropy in their model. 
However, in their recent work D12 break the degeneracy among their model parameters 
and measure $\beta = 0.4_{-0.2}^{+0.15}$ with $\alpha=4.6$ in the region
$16\lesssim r/\kpc \lesssim48$. Within the range of uncertainty our value for 
$\beta$ ($-0.14^{+0.52}_{-0.66}$) using GVE model agrees with them. 
We find that D12 value of $\beta\approx0.4$ in this bin, 
although derive from a sample which is dominated by stars within $r<27$ kpc,
is not appropriate for the range $18<r/\kpc<23$, instead it is appropriate
for the range $23<r/\kpc<48$. 
Finally, we check how well we can estimate the mass and anisotropy together using D11 DF
in the outer-most region ($35<r/\kpc<84$) 
We find that due to the lack of tangential information, a degeneracy
between mass-anisotropy cannot be broken.

Substituting the estimates of $\rsigma(r)$ and $\beta(r)$ in the Jeans equation, 
we then calculate the circular velocity profile of the Galaxy ($\vcirc(r)$). 
We detect the dip in the $\vcirc$ profile at 10-12 kpc, also seen by \cite{2009PASJ...61..227S} at 9 kpc, 
which is attributed to the massless ring as a perturbation to the disk. 
Finally, we fit the three component (exponential disk, Hernquist bulge, and NFW halo) 
galaxy model to the observed $\vcirc$ profile in order to obtain the mass distribution of the Galaxy. 
From our best fit model, we calculate $M_{\rm vir}$ of the halo to be 
$0.9^{+0.4}_{-0.3} \times 10^{12}$ $\msun$ with $R_{\rm vir} = 249^{+34}_{-31}\:\kpc$ 
and concentration parameter, $c = 12.0^{+0.6}_{-0.5}$. The mass of the Galaxy, within the 
extent we are able to constrain all the three components of velocity
dispersions, is estimated to be $M(r \lesssim 25 \:\kpc) = 2.1 \times 10^{11}$ $\msun$. Our estimate for 
$M_{\rm vir}$ is in good agreement with the most of the recent estimates
namely by \cite{2010MNRAS.406..264W, 1996ApJ...457..228K, 1999MNRAS.310..645W,  
2003A&A...397..899S, 2009PASJ...61..227S, 2010MNRAS.406..264W, 2012MNRAS.425.2840D} 
as demonstrated in Figure \ref{fig:vcirc_all}.  
In their studies of same population of stars (BHB), \cite{2008ApJ...684.1143X} also fit a three component
galaxy model and calculates $M_{\rm vir}$ to be $0.91^{+0.27}_{-0.18}\times 10^{12}$ $\msun$.
Our result for $M_{\rm vir}$ is in very good agreement with their estimate 
but our uncertainty is slightly larger.
Note that they make an assumption about the ($\vcirc/\vlos$) with radius
from the simulations whereas we do not make any such assumption.
Additionally, we also consider a more realistic three-dimensional 
disk model which is found to predict slightly higher Galactic mass 
$M_{\rm vir} = 1.2 ^{+0.5}_{-0.4}\times 10^{12}$ $\msun$ with
$R_{\rm vir} = 274^{+35}_{-30}\:\kpc$ for flattening constant $a = 2.5$.

In the end, we used the measured quantities $\rsigma$ and $\vcirc$ to extend the $\beta$ profile
beyond $r\sim25$ kpc up to the distance where $\rsigma(r)$ can be confidently measured ($r\sim60$ kpc).
The only assumption that we make here is about the density profile which we choose 
to be $\alpha = 4.5$ in consent to the recent results by \cite{2009MNRAS.398.1757W,2011MNRAS.416.2903D}. 
We find that the outer halo is radial and attains $\beta = 0.55$ at $r\sim60$ kpc.

We also compare our result with simulated stellar halo which
are formed purely by accretion \citep{2005ApJ...635..931B}.
This simulated halo is found to be in rough agreement to the observed halo
in the inner region $r<12$ kpc. 
It is seen that in none of the instances of simulations the $\beta$ profiles obtained 
could predict tangential halo at any distance and thus 
fails to explain a dip seen at $r=17$ kpc in observed $\beta$ profile.
In contrast, in the outer region ($r>25$ kpc) simulations and observations
both agree in overall sense of the anisotropy and predict a flat 
anisotropy profile.

In all the observed quantities $\rsigma, \tsigma, \psigma$ and $\beta$ we see
a dramatic shift in properties at $r\sim17$ kpc. 
We noticed that these undulations in the profiles are translated into our $\vcirc$ estimation
resulting a varying $\vcirc$ profile. 
It could be true other way around, in a sense that the non-monotonic trends seen in all 
of our kinematic profiles could be due to the presence of so far 
unaccounted features in the Milky Way potential.

Alternatively, the shift in the properties seen in the observed profiles 
could possibly be an indication of a complex multi-component halo.   
Recently there have been series of works advocating multi-component halo.
The studies of the calibration stars by \citet{2007Natur.450.1020C} and 
\citet{2010ApJ...712..692C} from the SDSS survey and the follow-up 
studies by \citet{2012ApJ...746...34B} have shown that the halo has at-least two 
distinct components. They associated the inner-halo to be formed in-situ
whereas the outer halo are considered to be formed by accretion.
\citet{2010ApJ...712..692C} and \citet{2010ApJ...714..663D} have  
found that the population fraction inversion point between the inner
and outer halo lies between $\sim15-20$ kpc. 
\citet{2012MNRAS.422.2116K} studied the BHB and RR Lyrae stars towards the galactic-anticenter and 
North Galactic Pole and found that the retrograde component of the halo dominates for $r>12.5$ kpc. 
It seems that this transition between the inner to the outer halo
is recorded in the $\beta$ of the halo as well. 
Additionally, duality in the formation history of the halo has also been seen 
in the recent smooth particle hydrodynamics and N-body simulations by 
\citet{2009ApJ...702.1058Z, 2012MNRAS.420.2245M, 2011MNRAS.417.1260F}. 
\citet{2012Natur.486...90K} recently attribute the age difference of 2 billion years in the halo components
to in-situ and accretion. 
On the contrary, \citet{2011MNRAS.415.3807S} reanalyzed the calibration stars
from \citet{2010ApJ...712..692C} and find no reliable evidence of the 
existence of outer retrograde halo.

In a nutshell, the stellar halo is a test bed to understand the formation
history of the galaxy. 
Even in this era where we have access to huge volume of spectroscopic and photometric data,
the crucial physical quantities like velocity dispersions and anisotropy are not
completely understood due to the lack of proper motions.
With the advent of data inflowing in the coming decades 
through the magnificent next generation of spectroscopic survey 
like LEGUE \citep{2012RAA....12..735D} and specially, unprecedented proper motions from
an astrometric mission like GAIA \citep{2002Ap&SS.280....1P} will help to put strong constrains on these 
fundamental quantities. 
Additionally, to see a bigger picture, confirming the results with the different stellar 
types or an alternative tracers is also crucial. 
Moreover, exploring the southern sky is equally important to complete 
the picture, for which up-coming spectroscopic survey like GALAH 
\footnote{\url{http://www.aao.gov.au/HERMES/GALAH/Home.html}} will also play an important role.

\section*{ACKNOWLEDGEMENT}
We sincerely thank the anonymous referee for comments those helped to improve the 
paper. We also thank Dr. Ralph Schonrich and Francesco Fermani
for their comments on the manuscript, particularly on the effect of 
(v$_{\rm LSR}$, R$_{\sun}$). 
Dr. Xiang Xiang Xue is thanked for the on-line publication of the clean sample of BHB stars. 
Mr. Tim White is also thanked for the comments on original manuscript.
PRK acknowledges University of Sydney International Scholarship (USydIS) for 
the support of his candidature. GFL acknowledges support from
ARC Discovery Project DP0665574. J.B.H. is funded through a Federation Fellowship from the
Australian Research Council (ARC) and
S.S. is funded through ARC DP grant 0988751 which supports the HERMES project.

Funding for the SDSS and SDSS-II has been provided by the Alfred P. Sloan Foundation,
the Participating Institutions, the National Science Foundation, the U.S. Department of Energy,
the National Aeronautics and Space Administration, the Japanese Monbukagakusho,
the Max Planck Society, and the Higher Education Funding Council for England. 
The SDSS Web Site is \url{http://www.sdss.org/.}

\appendix
\section{A. Binning and Effect of the bin size}
\label{sec:binning}
\begin{figure}
    \includegraphics[width=0.95\textwidth]{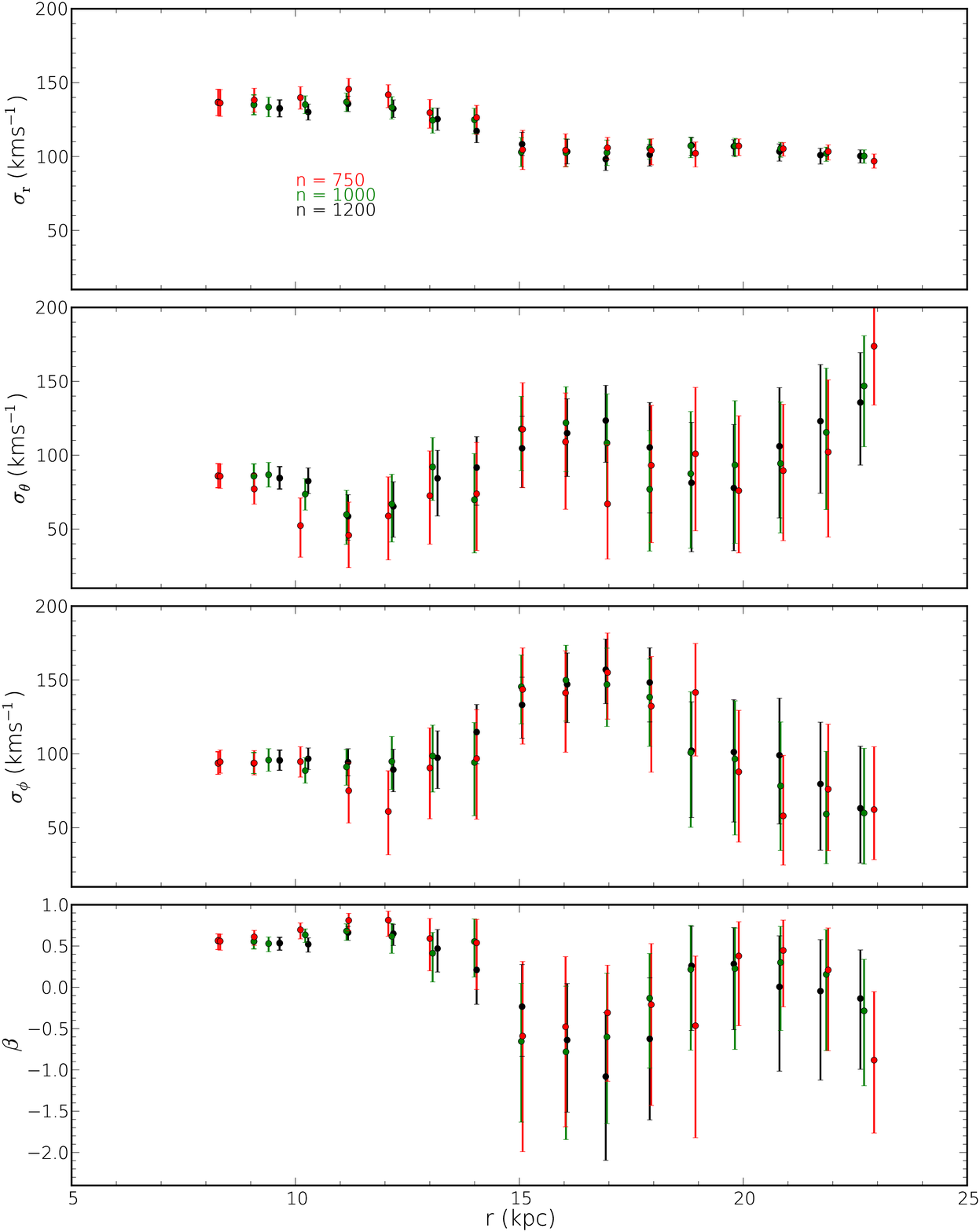}
  \caption{Effect of the number of particles in each CME bin. Red, green and 
           blue points are the estimates with 750, 1000 and 1200 number of stars in each bin respectively.
           From top to bottom are the $\rsigma$, $\tsigma$, $\psigma$ and anisotropy profile.}
\label{fig:effect_of_the_binning}
\end{figure}

Here we investigate an effect of $n_{\rm bin}$ to our analysis. 
Figure \ref{fig:effect_of_the_binning} shows the velocity 
dispersion profiles and the anisotropy profile for the same sample of stars 
but with different particles in each bin. We see that with the decrease in the number
of stars in each bin the uncertainties in the result increases. But the
overall trend remains unaltered.

\section{B. Effect of \lowercase{v}$_{\rm LSR}$ and R$_\sun$}
\label{sec:vlsrrsun_effect}
\begin{figure}
    \includegraphics[width=0.95\textwidth]{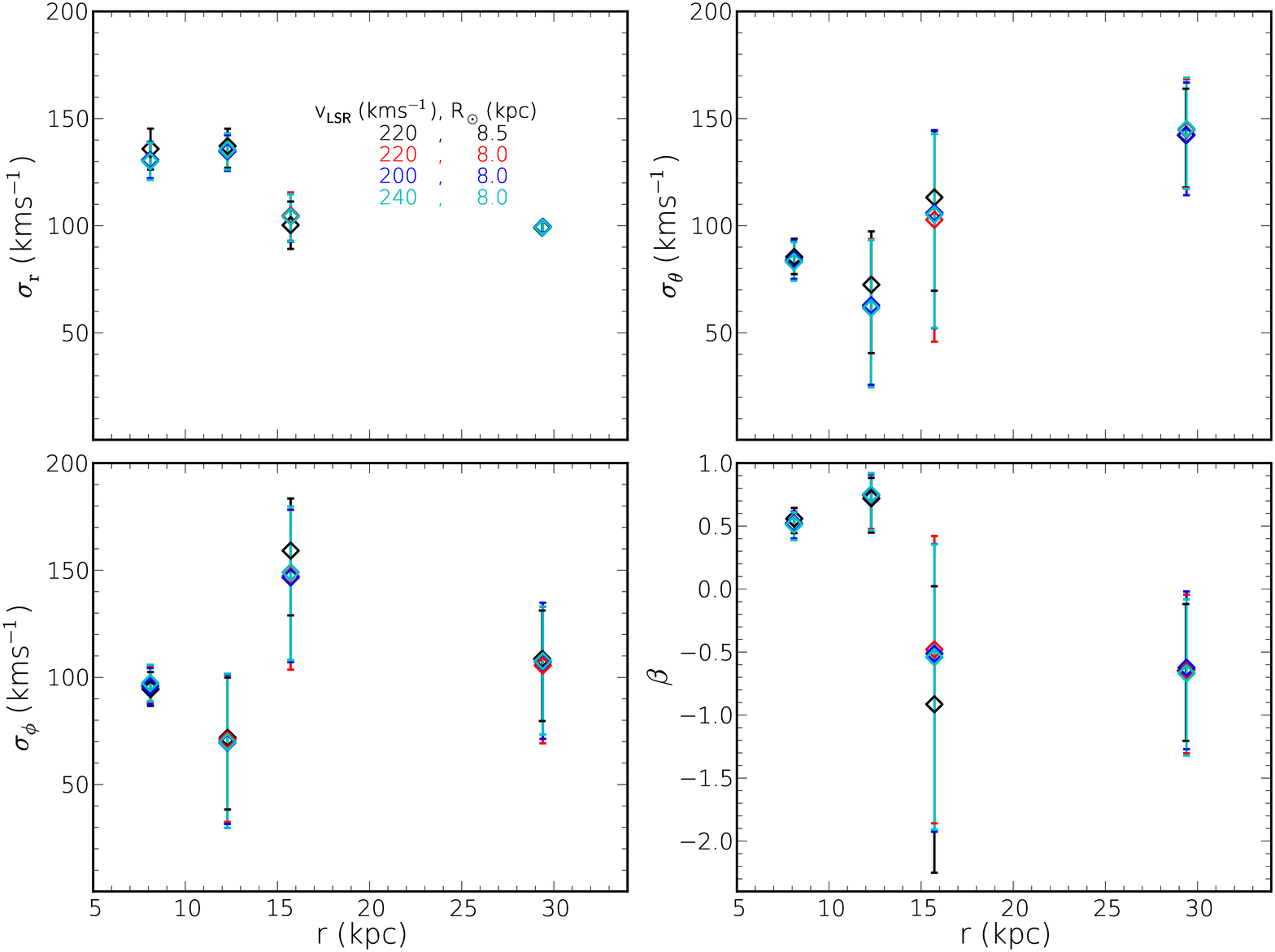}
  \caption{Velocity dispersion and anisotropy profiles for different  
           combinations of v$_{\rm LSR}$ and R$_\sun$.}
\label{fig:effect_vlsrRsun}
\end{figure}

In literatures there are varied claims about the value of v$_{\rm LSR}$ ranging from 
184--270 kms$^{-1}$ \citep{1998MNRAS.297..943O, 1999ApJ...524L..39M, 
2009ApJ...700..137R, 2009ApJ...704.1704B, 2010ApJ...712..260K}.
Similarly, the value of R$_\sun$ is also disputable
within 8--8.5 kpc \citep{1993ARA&A..31..345R, 2008ApJ...689.1044G, 2009ApJ...692.1075G}.
\cite{2010MNRAS.402..934M} found that the ratio v$_{\rm LSR}$/R$_\sun$ 
can be better constrained than each of them alone 
and should range between $29.9-31.6$ kms$^{-1}$kpc$^{-1}$.
Distressingly, there is still no consensus upon the values
of (v$_{\rm LSR}$, R$_\sun$). 
To study the effect of chosen values of (v$_{\rm LSR}$, R$_\sun$) 
upon our estimates of dispersion profiles
we repeat the same analysis done to obtain the 
black diamond points in \S \ref{sec:dispersion_profile} (Figure \ref{fig:aniso}).
In Figure \ref{fig:effect_vlsrRsun} we show the results
for different values of (v$_{\rm LSR}$, R$_\sun$). 
Here again the black diamond markers are 
obtained for a case (v$_{\rm LSR}$, R$_\sun$) = (220.0 kms$^{-1}$, 8.5 kpc) 
and is thus a replica of diamond points from Figure \ref{fig:aniso} 
put again for the ease of comparison. 
The red and black markers in the figure show the effect of chosen R$_{\sun}$ values
upon our estimates whereas the red, blue and cyan markers 
demonstrate the effect of chosen v$_{\rm LSR}$. 
Within the range of (v$_{\rm LSR}$, R$_\sun$) investigated, the figure clearly 
demonstrates a negligible effect of them upon our estimates, given the range
of uncertainties. 

\newpage
\bibliographystyle{apj}
\bibliography{\path/stellar_halo,\path/dispersion_profile,\path/anisotropic_model,\path/misc_reference,\path/mass_estimator,\path/reference_veldisp_paper,\path/msto}

\begin{thebibliography}{114}
\expandafter\ifx\csname natexlab\endcsname\relax\def\natexlab#1{#1}\fi

\bibitem[{{Abadi} {et~al.}(2006){Abadi}, {Navarro}, \&
  {Steinmetz}}]{2006MNRAS.365..747A}
{Abadi}, M.~G., {Navarro}, J.~F., \& {Steinmetz}, M. 2006, \mnras, 365, 747

\bibitem[{{Abazajian} {et~al.}(2009){Abazajian}, {Adelman-McCarthy},
  {Ag{\"u}eros}, {Allam}, {Allende Prieto}, {An}, {Anderson}, {Anderson},
  {Annis}, {Bahcall}, \& et~al.}]{2009ApJS..182..543A}
{Abazajian}, K.~N., {Adelman-McCarthy}, J.~K., {Ag{\"u}eros}, M.~A., {et~al.}
  2009, \apjs, 182, 543

\bibitem[{{Aihara} {et~al.}(2011){Aihara}, {Allende Prieto}, {An}, {Anderson},
  {Aubourg}, {Balbinot}, {Beers}, {Berlind}, {Bickerton}, {Bizyaev}, {Blanton},
  {Bochanski}, {Bolton}, {Bovy}, {Brandt}, {Brinkmann}, {Brown}, {Brownstein},
  {Busca}, {Campbell}, {Carr}, {Chen}, {Chiappini}, {Comparat}, {Connolly},
  {Cortes}, {Croft}, {Cuesta}, {da Costa}, {Davenport}, {Dawson}, {Dhital},
  {Ealet}, {Ebelke}, {Edmondson}, {Eisenstein}, {Escoffier}, {Esposito},
  {Evans}, {Fan}, {Femen{\'{\i}}a Castell{\'a}}, {Font-Ribera}, {Frinchaboy},
  {Ge}, {Gillespie}, {Gilmore}, {Gonz{\'a}lez Hern{\'a}ndez}, {Gott}, {Gould},
  {Grebel}, {Gunn}, {Hamilton}, {Harding}, {Harris}, {Hawley}, {Hearty}, {Ho},
  {Hogg}, {Holtzman}, {Honscheid}, {Inada}, {Ivans}, {Jiang}, {Johnson},
  {Jordan}, {Jordan}, {Kazin}, {Kirkby}, {Klaene}, {Knapp}, {Kneib},
  {Kochanek}, {Koesterke}, {Kollmeier}, {Kron}, {Lampeitl}, {Lang}, {Le Goff},
  {Lee}, {Lin}, {Long}, {Loomis}, {Lucatello}, {Lundgren}, {Lupton}, {Ma},
  {MacDonald}, {Mahadevan}, {Maia}, {Makler}, {Malanushenko}, {Malanushenko},
  {Mandelbaum}, {Maraston}, {Margala}, {Masters}, {McBride}, {McGehee},
  {McGreer}, {M{\'e}nard}, {Miralda-Escud{\'e}}, {Morrison}, {Mullally},
  {Muna}, {Munn}, {Murayama}, {Myers}, {Naugle}, {Neto}, {Nguyen}, {Nichol},
  {O'Connell}, {Ogando}, {Olmstead}, {Oravetz}, {Padmanabhan},
  {Palanque-Delabrouille}, {Pan}, {Pandey}, {P{\^a}ris}, {Percival},
  {Petitjean}, {Pfaffenberger}, {Pforr}, {Phleps}, {Pichon}, {Pieri}, {Prada},
  {Price-Whelan}, {Raddick}, {Ramos}, {Reyl{\'e}}, {Rich}, {Richards}, {Rix},
  {Robin}, {Rocha-Pinto}, {Rockosi}, {Roe}, {Rollinde}, {Ross}, {Ross},
  {Rossetto}, {S{\'a}nchez}, {Sayres}, {Schlegel}, {Schlesinger}, {Schmidt},
  {Schneider}, {Sheldon}, {Shu}, {Simmerer}, {Simmons}, {Sivarani}, {Snedden},
  {Sobeck}, {Steinmetz}, {Strauss}, {Szalay}, {Tanaka}, {Thakar}, {Thomas},
  {Tinker}, {Tofflemire}, {Tojeiro}, {Tremonti}, {Vandenberg}, {Vargas
  Maga{\~n}a}, {Verde}, {Vogt}, {Wake}, {Wang}, {Weaver}, {Weinberg}, {White},
  {White}, {Yanny}, {Yasuda}, {Yeche}, \& {Zehavi}}]{2011ApJS..193...29A}
{Aihara}, H., {Allende Prieto}, C., {An}, D., {et~al.} 2011, \apjs, 193, 29

\bibitem[{{Baes} \& {Dejonghe}(2002)}]{2002A&A...393..485B}
{Baes}, M., \& {Dejonghe}, H. 2002, \aap, 393, 485

\bibitem[{{Baes} \& {van Hese}(2007)}]{2007A&A...471..419B}
{Baes}, M., \& {van Hese}, E. 2007, \aap, 471, 419

\bibitem[{{Baiesi Pillastrini}(2009)}]{2009MNRAS.397.1990B}
{Baiesi Pillastrini}, G.~C. 2009, \mnras, 397, 1990

\bibitem[{{Battaglia} {et~al.}(2005){Battaglia}, {Helmi}, {Morrison},
  {Harding}, {Olszewski}, {Mateo}, {Freeman}, {Norris}, \&
  {Shectman}}]{2005MNRAS.364..433B}
{Battaglia}, G., {Helmi}, A., {Morrison}, H., {et~al.} 2005, \mnras, 364, 433

\bibitem[{{Beers} {et~al.}(2012){Beers}, {Carollo}, {Ivezi{\'c}}, {An},
  {Chiba}, {Norris}, {Freeman}, {Lee}, {Munn}, {Re Fiorentin}, {Sivarani},
  {Wilhelm}, {Yanny}, \& {York}}]{2012ApJ...746...34B}
{Beers}, T.~C., {Carollo}, D., {Ivezi{\'c}}, {\v Z}., {et~al.} 2012, \apj, 746,
  34

\bibitem[{{Bell} {et~al.}(2008){Bell}, {Zucker}, {Belokurov}, {Sharma},
  {Johnston}, {Bullock}, {Hogg}, {Jahnke}, {de Jong}, {Beers}, {Evans},
  {Grebel}, {Ivezi{\'c}}, {Koposov}, {Rix}, {Schneider}, {Steinmetz}, \&
  {Zolotov}}]{2008ApJ...680..295B}
{Bell}, E.~F., {Zucker}, D.~B., {Belokurov}, V., {et~al.} 2008, \apj, 680, 295

\bibitem[{{Belokurov} {et~al.}(2006){Belokurov}, {Zucker}, {Evans}, {Gilmore},
  {Vidrih}, {Bramich}, {Newberg}, {Wyse}, {Irwin}, {Fellhauer}, {Hewett},
  {Walton}, {Wilkinson}, {Cole}, {Yanny}, {Rockosi}, {Beers}, {Bell},
  {Brinkmann}, {Ivezi{\'c}}, \& {Lupton}}]{2006ApJ...642L.137B}
{Belokurov}, V., {Zucker}, D.~B., {Evans}, N.~W., {et~al.} 2006, \apjl, 642,
  L137

\bibitem[{{Besla} {et~al.}(2007){Besla}, {Kallivayalil}, {Hernquist},
  {Robertson}, {Cox}, {van der Marel}, \& {Alcock}}]{2007ApJ...668..949B}
{Besla}, G., {Kallivayalil}, N., {Hernquist}, L., {et~al.} 2007, \apj, 668, 949

\bibitem[{{Binney} \& {Tremaine}(2008)}]{2008gady.book.....B}
{Binney}, J., \& {Tremaine}, S. 2008, {Galactic Dynamics: Second Edition}
  (Princeton University Press)

\bibitem[{{Bland-Hawthorn}(2009)}]{2009IAUS..254..241B}
{Bland-Hawthorn}, J. 2009, in IAU Symposium, Vol. 254, IAU Symposium, ed.
  J.~{Andersen}, J.~{Bland-Hawthorn}, \& B.~{Nordstr{\"o}m}, 241--254

\bibitem[{{Blitz} {et~al.}(1982){Blitz}, {Fich}, \&
  {Stark}}]{1982ApJS...49..183B}
{Blitz}, L., {Fich}, M., \& {Stark}, A.~A. 1982, \apjs, 49, 183

\bibitem[{{Bond} {et~al.}(2010){Bond}, {Ivezi{\'c}}, {Sesar}, {Juri{\'c}},
  {Munn}, {Kowalski}, {Loebman}, {Ro{\v s}kar}, {Beers}, {Dalcanton},
  {Rockosi}, {Yanny}, {Newberg}, {Allende Prieto}, {Wilhelm}, {Lee},
  {Sivarani}, {Majewski}, {Norris}, {Bailer-Jones}, {Re Fiorentin}, {Schlegel},
  {Uomoto}, {Lupton}, {Knapp}, {Gunn}, {Covey}, {Allyn Smith}, {Miknaitis},
  {Doi}, {Tanaka}, {Fukugita}, {Kent}, {Finkbeiner}, {Quinn}, {Hawley},
  {Anderson}, {Kiuchi}, {Chen}, {Bushong}, {Sohi}, {Haggard}, {Kimball},
  {McGurk}, {Barentine}, {Brewington}, {Harvanek}, {Kleinman}, {Krzesinski},
  {Long}, {Nitta}, {Snedden}, {Lee}, {Pier}, {Harris}, {Brinkmann}, \&
  {Schneider}}]{2010ApJ...716....1B}
{Bond}, N.~A., {Ivezi{\'c}}, {\v Z}., {Sesar}, B., {et~al.} 2010, \apj, 716, 1

\bibitem[{{Bovy} {et~al.}(2009){Bovy}, {Hogg}, \& {Rix}}]{2009ApJ...704.1704B}
{Bovy}, J., {Hogg}, D.~W., \& {Rix}, H.-W. 2009, \apj, 704, 1704

\bibitem[{{Brown} {et~al.}(2010){Brown}, {Geller}, {Kenyon}, \&
  {Diaferio}}]{2010AJ....139...59B}
{Brown}, W.~R., {Geller}, M.~J., {Kenyon}, S.~J., \& {Diaferio}, A. 2010, \aj,
  139, 59

\bibitem[{{Bryan} \& {Norman}(1998)}]{1998ApJ...495...80B}
{Bryan}, G.~L., \& {Norman}, M.~L. 1998, \apj, 495, 80

\bibitem[{{Bullock} \& {Johnston}(2005)}]{2005ApJ...635..931B}
{Bullock}, J.~S., \& {Johnston}, K.~V. 2005, \apj, 635, 931

\bibitem[{{Burton} \& {Gordon}(1978)}]{1978A&A....63....7B}
{Burton}, W.~B., \& {Gordon}, M.~A. 1978, \aap, 63, 7

\bibitem[{{Carollo} {et~al.}(2007){Carollo}, {Beers}, {Lee}, {Chiba}, {Norris},
  {Wilhelm}, {Sivarani}, {Marsteller}, {Munn}, {Bailer-Jones}, {Fiorentin}, \&
  {York}}]{2007Natur.450.1020C}
{Carollo}, D., {Beers}, T.~C., {Lee}, Y.~S., {et~al.} 2007, \nat, 450, 1020

\bibitem[{{Carollo} {et~al.}(2010){Carollo}, {Beers}, {Chiba}, {Norris},
  {Freeman}, {Lee}, {Ivezi{\'c}}, {Rockosi}, \& {Yanny}}]{2010ApJ...712..692C}
{Carollo}, D., {Beers}, T.~C., {Chiba}, M., {et~al.} 2010, \apj, 712, 692

\bibitem[{{Clemens}(1985)}]{1985ApJ...295..422C}
{Clemens}, D.~P. 1985, \apj, 295, 422

\bibitem[{{Cuddeford}(1991)}]{1991MNRAS.253..414C}
{Cuddeford}, P. 1991, \mnras, 253, 414

\bibitem[{{de Jong} {et~al.}(2010){de Jong}, {Yanny}, {Rix}, {Dolphin},
  {Martin}, \& {Beers}}]{2010ApJ...714..663D}
{de Jong}, J.~T.~A., {Yanny}, B., {Rix}, H.-W., {et~al.} 2010, \apj, 714, 663

\bibitem[{{De Propris} {et~al.}(2010){De Propris}, {Harrison}, \&
  {Mares}}]{2010ApJ...719.1582D}
{De Propris}, R., {Harrison}, C.~D., \& {Mares}, P.~J. 2010, \apj, 719, 1582

\bibitem[{{Deason} {et~al.}(2011{\natexlab{a}}){Deason}, {Belokurov}, \&
  {Evans}}]{2011MNRAS.411.1480D}
{Deason}, A.~J., {Belokurov}, V., \& {Evans}, N.~W. 2011{\natexlab{a}}, \mnras,
  411, 1480

\bibitem[{{Deason} {et~al.}(2011{\natexlab{b}}){Deason}, {Belokurov}, \&
  {Evans}}]{2011MNRAS.416.2903D}
---. 2011{\natexlab{b}}, \mnras, 416, 2903

\bibitem[{{Deason} {et~al.}(2012{\natexlab{a}}){Deason}, {Belokurov}, {Evans},
  \& {An}}]{2012MNRAS.tmpL.469D}
{Deason}, A.~J., {Belokurov}, V., {Evans}, N.~W., \& {An}, J.
  2012{\natexlab{a}}, \mnras, L469

\bibitem[{{Deason} {et~al.}(2012{\natexlab{b}}){Deason}, {Belokurov}, {Evans},
  {Koposov}, {Cooke}, {Pe{\~n}arrubia}, {Laporte}, {Fellhauer}, {Walker}, \&
  {Olszewski}}]{2012MNRAS.425.2840D}
{Deason}, A.~J., {Belokurov}, V., {Evans}, N.~W., {et~al.} 2012{\natexlab{b}},
  \mnras, 425, 2840

\bibitem[{{Dehnen} {et~al.}(2006){Dehnen}, {McLaughlin}, \&
  {Sachania}}]{2006MNRAS.369.1688D}
{Dehnen}, W., {McLaughlin}, D.~E., \& {Sachania}, J. 2006, \mnras, 369, 1688

\bibitem[{{Demers} \& {Battinelli}(2007)}]{2007A&A...473..143D}
{Demers}, S., \& {Battinelli}, P. 2007, \aap, 473, 143

\bibitem[{{Deng} {et~al.}(2012){Deng}, {Newberg}, {Liu}, {Carlin}, {Beers},
  {Chen}, {Chen}, {Christlieb}, {Grillmair}, {Guhathakurta}, {Han}, {Hou},
  {Lee}, {L{\'e}pine}, {Li}, {Liu}, {Pan}, {Sellwood}, {Wang}, {Wang}, {Yang},
  {Yanny}, {Zhang}, {Zhang}, {Zheng}, \& {Zhu}}]{2012RAA....12..735D}
{Deng}, L.-C., {Newberg}, H.~J., {Liu}, C., {et~al.} 2012, Research in
  Astronomy and Astrophysics, 12, 735

\bibitem[{{Eisenhauer} {et~al.}(2005){Eisenhauer}, {Genzel}, {Alexander},
  {Abuter}, {Paumard}, {Ott}, {Gilbert}, {Gillessen}, {Horrobin}, {Trippe},
  {Bonnet}, {Dumas}, {Hubin}, {Kaufer}, {Kissler-Patig}, {Monnet},
  {Str{\"o}bele}, {Szeifert}, {Eckart}, {Sch{\"o}del}, \&
  {Zucker}}]{2005ApJ...628..246E}
{Eisenhauer}, F., {Genzel}, R., {Alexander}, T., {et~al.} 2005, \apj, 628, 246

\bibitem[{{Eneev} {et~al.}(1973){Eneev}, {Kozlov}, \&
  {Sunyaev}}]{1973A&A....22...41E}
{Eneev}, T.~M., {Kozlov}, N.~N., \& {Sunyaev}, R.~A. 1973, \aap, 22, 41

\bibitem[{{Evans} {et~al.}(1997){Evans}, {Hafner}, \& {de
  Zeeuw}}]{1997MNRAS.286..315E}
{Evans}, N.~W., {Hafner}, R.~M., \& {de Zeeuw}, P.~T. 1997, \mnras, 286, 315

\bibitem[{{Fich} {et~al.}(1989){Fich}, {Blitz}, \&
  {Stark}}]{1989ApJ...342..272F}
{Fich}, M., {Blitz}, L., \& {Stark}, A.~A. 1989, \apj, 342, 272

\bibitem[{{Font} {et~al.}(2011){Font}, {Benson}, {Bower}, {Frenk}, {Cooper},
  {De Lucia}, {Helly}, {Helmi}, {Li}, {McCarthy}, {Navarro}, {Springel},
  {Starkenburg}, {Wang}, \& {White}}]{2011MNRAS.417.1260F}
{Font}, A.~S., {Benson}, A.~J., {Bower}, R.~G., {et~al.} 2011, \mnras, 417,
  1260

\bibitem[{{Freeman} \& {Bland-Hawthorn}(2002)}]{2002ARA&A..40..487F}
{Freeman}, K., \& {Bland-Hawthorn}, J. 2002, \araa, 40, 487

\bibitem[{{Freeman}(1970)}]{1970ApJ...160..811F}
{Freeman}, K.~C. 1970, \apj, 160, 811

\bibitem[{{Frenk} \& {White}(1980)}]{1980MNRAS.193..295F}
{Frenk}, C.~S., \& {White}, S.~D.~M. 1980, \mnras, 193, 295

\bibitem[{{Gerhard}(1991)}]{1991MNRAS.250..812G}
{Gerhard}, O.~E. 1991, \mnras, 250, 812

\bibitem[{{Ghez} {et~al.}(2008){Ghez}, {Salim}, {Weinberg}, {Lu}, {Do}, {Dunn},
  {Matthews}, {Morris}, {Yelda}, {Becklin}, {Kremenek}, {Milosavljevic}, \&
  {Naiman}}]{2008ApJ...689.1044G}
{Ghez}, A.~M., {Salim}, S., {Weinberg}, N.~N., {et~al.} 2008, \apj, 689, 1044

\bibitem[{{Gillessen} {et~al.}(2009){Gillessen}, {Eisenhauer}, {Trippe},
  {Alexander}, {Genzel}, {Martins}, \& {Ott}}]{2009ApJ...692.1075G}
{Gillessen}, S., {Eisenhauer}, F., {Trippe}, S., {et~al.} 2009, \apj, 692, 1075

\bibitem[{{Gnedin} {et~al.}(2010){Gnedin}, {Brown}, {Geller}, \&
  {Kenyon}}]{2010ApJ...720L.108G}
{Gnedin}, O.~Y., {Brown}, W.~R., {Geller}, M.~J., \& {Kenyon}, S.~J. 2010,
  \apjl, 720, L108

\bibitem[{{Helmi}(2008)}]{2008A&ARv..15..145H}
{Helmi}, A. 2008, \aapr, 15, 145

\bibitem[{{Helmi} \& {White}(1999)}]{1999MNRAS.307..495H}
{Helmi}, A., \& {White}, S.~D.~M. 1999, \mnras, 307, 495

\bibitem[{{Henon}(1973)}]{1973A&A....24..229H}
{Henon}, M. 1973, \aap, 24, 229

\bibitem[{{Honma} \& {Sofue}(1997{\natexlab{a}})}]{1997PASJ...49..539H}
{Honma}, M., \& {Sofue}, Y. 1997{\natexlab{a}}, \pasj, 49, 539

\bibitem[{{Honma} \& {Sofue}(1997{\natexlab{b}})}]{1997PASJ...49..453H}
---. 1997{\natexlab{b}}, \pasj, 49, 453

\bibitem[{{Honma} {et~al.}(2007){Honma}, {Bushimata}, {Choi}, {Hirota}, {Imai},
  {Iwadate}, {Jike}, {Kameya}, {Kamohara}, {Kan-Ya}, {Kawaguchi}, {Kijima},
  {Kobayashi}, {Kuji}, {Kurayama}, {Manabe}, {Miyaji}, {Nagayama}, {Nakagawa},
  {Oh}, {Omodaka}, {Oyama}, {Sakai}, {Sato}, {Sasao}, {Shibata}, {Shintani},
  {Suda}, {Tamura}, {Tsushima}, \& {Yamashita Kazuyoshi}}]{2007PASJ...59..889H}
{Honma}, M., {Bushimata}, T., {Choi}, Y.~K., {et~al.} 2007, \pasj, 59, 889

\bibitem[{{Ibata} {et~al.}(1994){Ibata}, {Gilmore}, \&
  {Irwin}}]{1994Natur.370..194I}
{Ibata}, R.~A., {Gilmore}, G., \& {Irwin}, M.~J. 1994, \nat, 370, 194

\bibitem[{{Ibata} {et~al.}(1995){Ibata}, {Gilmore}, \&
  {Irwin}}]{1995MNRAS.277..781I}
---. 1995, \mnras, 277, 781

\bibitem[{{Jeans}(1915)}]{1915MNRAS..76...70J}
{Jeans}, J.~H. 1915, \mnras, 76, 70

\bibitem[{{Johnston} {et~al.}(2008){Johnston}, {Bullock}, {Sharma}, {Font},
  {Robertson}, \& {Leitner}}]{2008ApJ...689..936J}
{Johnston}, K.~V., {Bullock}, J.~S., {Sharma}, S., {et~al.} 2008, \apj, 689,
  936

\bibitem[{{Juri{\'c}} {et~al.}(2008){Juri{\'c}}, {Ivezi{\'c}}, {Brooks},
  {Lupton}, {Schlegel}, {Finkbeiner}, {Padmanabhan}, {Bond}, {Sesar},
  {Rockosi}, {Knapp}, {Gunn}, {Sumi}, {Schneider}, {Barentine}, {Brewington},
  {Brinkmann}, {Fukugita}, {Harvanek}, {Kleinman}, {Krzesinski}, {Long},
  {Neilsen}, {Nitta}, {Snedden}, \& {York}}]{2008ApJ...673..864J}
{Juri{\'c}}, M., {Ivezi{\'c}}, {\v Z}., {Brooks}, A., {et~al.} 2008, \apj, 673,
  864

\bibitem[{{Kalirai}(2012)}]{2012Natur.486...90K}
{Kalirai}, J.~S. 2012, \nat, 486, 90

\bibitem[{{Karachentsev} \& {Kashibadze}(2006)}]{2006Ap.....49....3K}
{Karachentsev}, I.~D., \& {Kashibadze}, O.~G. 2006, Astrophysics, 49, 3

\bibitem[{{Kinman} {et~al.}(2007){Kinman}, {Cacciari}, {Bragaglia}, {Buzzoni},
  \& {Spagna}}]{2007MNRAS.375.1381K}
{Kinman}, T.~D., {Cacciari}, C., {Bragaglia}, A., {Buzzoni}, A., \& {Spagna},
  A. 2007, \mnras, 375, 1381

\bibitem[{{Kinman} {et~al.}(2012){Kinman}, {Cacciari}, {Bragaglia}, {Smart}, \&
  {Spagna}}]{2012MNRAS.422.2116K}
{Kinman}, T.~D., {Cacciari}, C., {Bragaglia}, A., {Smart}, R., \& {Spagna}, A.
  2012, \mnras, 422, 2116

\bibitem[{{Kochanek}(1996)}]{1996ApJ...457..228K}
{Kochanek}, C.~S. 1996, \apj, 457, 228

\bibitem[{{Komatsu} {et~al.}(2011){Komatsu}, {Smith}, {Dunkley}, {Bennett},
  {Gold}, {Hinshaw}, {Jarosik}, {Larson}, {Nolta}, {Page}, {Spergel},
  {Halpern}, {Hill}, {Kogut}, {Limon}, {Meyer}, {Odegard}, {Tucker}, {Weiland},
  {Wollack}, \& {Wright}}]{2011ApJS..192...18K}
{Komatsu}, E., {Smith}, K.~M., {Dunkley}, J., {et~al.} 2011, \apjs, 192, 18

\bibitem[{{Koposov} {et~al.}(2010){Koposov}, {Rix}, \&
  {Hogg}}]{2010ApJ...712..260K}
{Koposov}, S.~E., {Rix}, H.-W., \& {Hogg}, D.~W. 2010, \apj, 712, 260

\bibitem[{{Li} \& {White}(2008)}]{2008MNRAS.384.1459L}
{Li}, Y.-S., \& {White}, S.~D.~M. 2008, \mnras, 384, 1459

\bibitem[{{Lin} \& {Lynden-Bell}(1982)}]{1982MNRAS.198..707L}
{Lin}, D.~N.~C., \& {Lynden-Bell}, D. 1982, \mnras, 198, 707

\bibitem[{{Macci{\`o}} {et~al.}(2007){Macci{\`o}}, {Dutton}, {van den Bosch},
  {Moore}, {Potter}, \& {Stadel}}]{2007MNRAS.378...55M}
{Macci{\`o}}, A.~V., {Dutton}, A.~A., {van den Bosch}, F.~C., {et~al.} 2007,
  \mnras, 378, 55

\bibitem[{{Majewski}(1993)}]{1993ARA&A..31..575M}
{Majewski}, S.~R. 1993, \araa, 31, 575

\bibitem[{{Majewski} {et~al.}(2004){Majewski}, {Ostheimer}, {Rocha-Pinto},
  {Patterson}, {Guhathakurta}, \& {Reitzel}}]{2004ApJ...615..738M}
{Majewski}, S.~R., {Ostheimer}, J.~C., {Rocha-Pinto}, H.~J., {et~al.} 2004,
  \apj, 615, 738

\bibitem[{{Majewski} {et~al.}(2003){Majewski}, {Skrutskie}, {Weinberg}, \&
  {Ostheimer}}]{2003ApJ...599.1082M}
{Majewski}, S.~R., {Skrutskie}, M.~F., {Weinberg}, M.~D., \& {Ostheimer}, J.~C.
  2003, \apj, 599, 1082

\bibitem[{{Majewski} {et~al.}(2007){Majewski}, {Beaton}, {Patterson},
  {Kalirai}, {Geha}, {Mu{\~n}oz}, {Seigar}, {Guhathakurta}, {Gilbert}, {Rich},
  {Bullock}, \& {Reitzel}}]{2007ApJ...670L...9M}
{Majewski}, S.~R., {Beaton}, R.~L., {Patterson}, R.~J., {et~al.} 2007, \apjl,
  670, L9

\bibitem[{{Martin} {et~al.}(2007){Martin}, {Ibata}, \&
  {Irwin}}]{2007ApJ...668L.123M}
{Martin}, N.~F., {Ibata}, R.~A., \& {Irwin}, M. 2007, \apjl, 668, L123

\bibitem[{{McCarthy} {et~al.}(2012){McCarthy}, {Font}, {Crain}, {Deason},
  {Schaye}, \& {Theuns}}]{2012MNRAS.420.2245M}
{McCarthy}, I.~G., {Font}, A.~S., {Crain}, R.~A., {et~al.} 2012, \mnras, 420,
  2245

\bibitem[{{McConnachie} {et~al.}(2005){McConnachie}, {Irwin}, {Ferguson},
  {Ibata}, {Lewis}, \& {Tanvir}}]{2005MNRAS.356..979M}
{McConnachie}, A.~W., {Irwin}, M.~J., {Ferguson}, A.~M.~N., {et~al.} 2005,
  \mnras, 356, 979

\bibitem[{{McMillan} \& {Binney}(2010)}]{2010MNRAS.402..934M}
{McMillan}, P.~J., \& {Binney}, J.~J. 2010, \mnras, 402, 934

\bibitem[{{M{\'e}ndez} {et~al.}(1999){M{\'e}ndez}, {Platais}, {Girard},
  {Kozhurina-Platais}, \& {van Altena}}]{1999ApJ...524L..39M}
{M{\'e}ndez}, R.~A., {Platais}, I., {Girard}, T.~M., {Kozhurina-Platais}, V.,
  \& {van Altena}, W.~F. 1999, \apjl, 524, L39

\bibitem[{{Merritt}(1985{\natexlab{a}})}]{1985MNRAS.214P..25M}
{Merritt}, D. 1985{\natexlab{a}}, \mnras, 214, 25P

\bibitem[{{Merritt}(1985{\natexlab{b}})}]{1985AJ.....90.1027M}
---. 1985{\natexlab{b}}, \aj, 90, 1027

\bibitem[{{Miyamoto} \& {Nagai}(1975)}]{1975PASJ...27..533M}
{Miyamoto}, M., \& {Nagai}, R. 1975, \pasj, 27, 533

\bibitem[{{Navarro} {et~al.}(1996){Navarro}, {Frenk}, \&
  {White}}]{1996ApJ...462..563N}
{Navarro}, J.~F., {Frenk}, C.~S., \& {White}, S.~D.~M. 1996, \apj, 462, 563

\bibitem[{{Newberg} {et~al.}(2002){Newberg}, {Yanny}, {Rockosi}, {Grebel},
  {Rix}, {Brinkmann}, {Csabai}, {Hennessy}, {Hindsley}, {Ibata}, {Ivezi{\'c}},
  {Lamb}, {Nash}, {Odenkirchen}, {Rave}, {Schneider}, {Smith}, {Stolte}, \&
  {York}}]{2002ApJ...569..245N}
{Newberg}, H.~J., {Yanny}, B., {Rockosi}, C., {et~al.} 2002, \apj, 569, 245

\bibitem[{{Olling} \& {Merrifield}(1998)}]{1998MNRAS.297..943O}
{Olling}, R.~P., \& {Merrifield}, M.~R. 1998, \mnras, 297, 943

\bibitem[{{Osipkov}(1979)}]{1979SvAL....5...42O}
{Osipkov}, L.~P. 1979, Soviet Astronomy Letters, 5, 42

\bibitem[{{Perryman}(2002)}]{2002Ap&SS.280....1P}
{Perryman}, M.~A.~C. 2002, \apss, 280, 1

\bibitem[{{Reid}(1993)}]{1993ARA&A..31..345R}
{Reid}, M.~J. 1993, \araa, 31, 345

\bibitem[{{Reid} {et~al.}(2009){Reid}, {Menten}, {Zheng}, {Brunthaler},
  {Moscadelli}, {Xu}, {Zhang}, {Sato}, {Honma}, {Hirota}, {Hachisuka}, {Choi},
  {Moellenbrock}, \& {Bartkiewicz}}]{2009ApJ...700..137R}
{Reid}, M.~J., {Menten}, K.~M., {Zheng}, X.~W., {et~al.} 2009, \apj, 700, 137

\bibitem[{{Ribas} {et~al.}(2005){Ribas}, {Jordi}, {Vilardell}, {Fitzpatrick},
  {Hilditch}, \& {Guinan}}]{2005ApJ...635L..37R}
{Ribas}, I., {Jordi}, C., {Vilardell}, F., {et~al.} 2005, \apjl, 635, L37

\bibitem[{{Rocha-Pinto} {et~al.}(2004){Rocha-Pinto}, {Majewski}, {Skrutskie},
  {Crane}, \& {Patterson}}]{2004ApJ...615..732R}
{Rocha-Pinto}, H.~J., {Majewski}, S.~R., {Skrutskie}, M.~F., {Crane}, J.~D., \&
  {Patterson}, R.~J. 2004, \apj, 615, 732

\bibitem[{{Sakamoto} {et~al.}(2003){Sakamoto}, {Chiba}, \&
  {Beers}}]{2003A&A...397..899S}
{Sakamoto}, T., {Chiba}, M., \& {Beers}, T.~C. 2003, \aap, 397, 899

\bibitem[{{Sales} {et~al.}(2007){Sales}, {Navarro}, {Abadi}, \&
  {Steinmetz}}]{2007MNRAS.379.1464S}
{Sales}, L.~V., {Navarro}, J.~F., {Abadi}, M.~G., \& {Steinmetz}, M. 2007,
  \mnras, 379, 1464

\bibitem[{{Samurovi{\'c}} \& {Lalovi{\'c}}(2011)}]{2011A&A...531A..82S}
{Samurovi{\'c}}, S., \& {Lalovi{\'c}}, A. 2011, \aap, 531, A82

\bibitem[{{Sch{\"o}del} {et~al.}(2002){Sch{\"o}del}, {Ott}, {Genzel},
  {Hofmann}, {Lehnert}, {Eckart}, {Mouawad}, {Alexander}, {Reid}, {Lenzen},
  {Hartung}, {Lacombe}, {Rouan}, {Gendron}, {Rousset}, {Lagrange}, {Brandner},
  {Ageorges}, {Lidman}, {Moorwood}, {Spyromilio}, {Hubin}, \&
  {Menten}}]{2002Natur.419..694S}
{Sch{\"o}del}, R., {Ott}, T., {Genzel}, R., {et~al.} 2002, \nat, 419, 694

\bibitem[{{Sch{\"o}nrich} {et~al.}(2011){Sch{\"o}nrich}, {Asplund}, \&
  {Casagrande}}]{2011MNRAS.415.3807S}
{Sch{\"o}nrich}, R., {Asplund}, M., \& {Casagrande}, L. 2011, \mnras, 415, 3807

\bibitem[{{Sch{\"o}nrich} {et~al.}(2010){Sch{\"o}nrich}, {Binney}, \&
  {Dehnen}}]{2010MNRAS.403.1829S}
{Sch{\"o}nrich}, R., {Binney}, J., \& {Dehnen}, W. 2010, \mnras, 403, 1829

\bibitem[{{Searle} \& {Zinn}(1978)}]{1978ApJ...225..357S}
{Searle}, L., \& {Zinn}, R. 1978, \apj, 225, 357

\bibitem[{{Seigar} {et~al.}(2008){Seigar}, {Barth}, \&
  {Bullock}}]{2008MNRAS.389.1911S}
{Seigar}, M.~S., {Barth}, A.~J., \& {Bullock}, J.~S. 2008, \mnras, 389, 1911

\bibitem[{{Sharma} {et~al.}(2011{\natexlab{a}}){Sharma}, {Bland-Hawthorn},
  {Johnston}, \& {Binney}}]{2011ApJ...730....3S}
{Sharma}, S., {Bland-Hawthorn}, J., {Johnston}, K.~V., \& {Binney}, J.
  2011{\natexlab{a}}, \apj, 730, 3

\bibitem[{{Sharma} {et~al.}(2011{\natexlab{b}}){Sharma}, {Johnston},
  {Majewski}, {Bullock}, \& {Mu{\~n}oz}}]{2011ApJ...728..106S}
{Sharma}, S., {Johnston}, K.~V., {Majewski}, S.~R., {Bullock}, J., \&
  {Mu{\~n}oz}, R.~R. 2011{\natexlab{b}}, \apj, 728, 106

\bibitem[{{Sirko} {et~al.}(2004{\natexlab{a}}){Sirko}, {Goodman}, {Knapp},
  {Brinkmann}, {Ivezi{\'c}}, {Knerr}, {Schlegel}, {Schneider}, \&
  {York}}]{2004AJ....127..899S}
{Sirko}, E., {Goodman}, J., {Knapp}, G.~R., {et~al.} 2004{\natexlab{a}}, \aj,
  127, 899

\bibitem[{{Sirko} {et~al.}(2004{\natexlab{b}}){Sirko}, {Goodman}, {Knapp},
  {Brinkmann}, {Ivezi{\'c}}, {Knerr}, {Schlegel}, {Schneider}, \&
  {York}}]{2004AJ....127..914S}
---. 2004{\natexlab{b}}, \aj, 127, 914

\bibitem[{{Smith} {et~al.}(2009{\natexlab{a}}){Smith}, {Wyn Evans}, \&
  {An}}]{2009ApJ...698.1110S}
{Smith}, M.~C., {Wyn Evans}, N., \& {An}, J.~H. 2009{\natexlab{a}}, \apj, 698,
  1110

\bibitem[{{Smith} {et~al.}(2007){Smith}, {Ruchti}, {Helmi}, {Wyse},
  {Fulbright}, {Freeman}, {Navarro}, {Seabroke}, {Steinmetz}, {Williams},
  {Bienaym{\'e}}, {Binney}, {Bland-Hawthorn}, {Dehnen}, {Gibson}, {Gilmore},
  {Grebel}, {Munari}, {Parker}, {Scholz}, {Siebert}, {Watson}, \&
  {Zwitter}}]{2007MNRAS.379..755S}
{Smith}, M.~C., {Ruchti}, G.~R., {Helmi}, A., {et~al.} 2007, \mnras, 379, 755

\bibitem[{{Smith} {et~al.}(2009{\natexlab{b}}){Smith}, {Evans}, {Belokurov},
  {Hewett}, {Bramich}, {Gilmore}, {Irwin}, {Vidrih}, \&
  {Zucker}}]{2009MNRAS.399.1223S}
{Smith}, M.~C., {Evans}, N.~W., {Belokurov}, V., {et~al.} 2009{\natexlab{b}},
  \mnras, 399, 1223

\bibitem[{{Sofue} {et~al.}(2009){Sofue}, {Honma}, \&
  {Omodaka}}]{2009PASJ...61..227S}
{Sofue}, Y., {Honma}, M., \& {Omodaka}, T. 2009, \pasj, 61, 227

\bibitem[{{Sommer-Larsen} {et~al.}(1997){Sommer-Larsen}, {Beers}, {Flynn},
  {Wilhelm}, \& {Christensen}}]{1997ApJ...481..775S}
{Sommer-Larsen}, J., {Beers}, T.~C., {Flynn}, C., {Wilhelm}, R., \&
  {Christensen}, P.~R. 1997, \apj, 481, 775

\bibitem[{{Toomre}(1963)}]{1963ApJ...138..385T}
{Toomre}, A. 1963, \apj, 138, 385

\bibitem[{{Watkins} {et~al.}(2010){Watkins}, {Evans}, \&
  {An}}]{2010MNRAS.406..264W}
{Watkins}, L.~L., {Evans}, N.~W., \& {An}, J.~H. 2010, \mnras, 406, 264

\bibitem[{{Watkins} {et~al.}(2009){Watkins}, {Evans}, {Belokurov}, {Smith},
  {Hewett}, {Bramich}, {Gilmore}, {Irwin}, {Vidrih}, {Wyrzykowski}, \&
  {Zucker}}]{2009MNRAS.398.1757W}
{Watkins}, L.~L., {Evans}, N.~W., {Belokurov}, V., {et~al.} 2009, \mnras, 398,
  1757

\bibitem[{{White} \& {Rees}(1978)}]{1978MNRAS.183..341W}
{White}, S.~D.~M., \& {Rees}, M.~J. 1978, \mnras, 183, 341

\bibitem[{{Wilkinson} \& {Evans}(1999)}]{1999MNRAS.310..645W}
{Wilkinson}, M.~I., \& {Evans}, N.~W. 1999, \mnras, 310, 645

\bibitem[{{Wolfire} {et~al.}(1995){Wolfire}, {McKee}, {Hollenbach}, \&
  {Tielens}}]{1995ApJ...453..673W}
{Wolfire}, M.~G., {McKee}, C.~F., {Hollenbach}, D., \& {Tielens}, A.~G.~G.~M.
  1995, \apj, 453, 673

\bibitem[{{Xue} {et~al.}(2008){Xue}, {Rix}, {Zhao}, {Re Fiorentin}, {Naab},
  {Steinmetz}, {van den Bosch}, {Beers}, {Lee}, {Bell}, {Rockosi}, {Yanny},
  {Newberg}, {Wilhelm}, {Kang}, {Smith}, \& {Schneider}}]{2008ApJ...684.1143X}
{Xue}, X.~X., {Rix}, H.~W., {Zhao}, G., {et~al.} 2008, \apj, 684, 1143

\bibitem[{{Xue} {et~al.}(2011){Xue}, {Rix}, {Yanny}, {Beers}, {Bell}, {Zhao},
  {Bullock}, {Johnston}, {Morrison}, {Rockosi}, {Koposov}, {Kang}, {Liu},
  {Luo}, {Lee}, \& {Weaver}}]{2011ApJ...738...79X}
{Xue}, X.-X., {Rix}, H.-W., {Yanny}, B., {et~al.} 2011, \apj, 738, 79

\bibitem[{{Yanny} {et~al.}(2009){Yanny}, {Rockosi}, {Newberg}, {Knapp},
  {Adelman-McCarthy}, {Alcorn}, {Allam}, {Allende Prieto}, {An}, {Anderson},
  {Anderson}, {Bailer-Jones}, {Bastian}, {Beers}, {Bell}, {Belokurov},
  {Bizyaev}, {Blythe}, {Bochanski}, {Boroski}, {Brinchmann}, {Brinkmann},
  {Brewington}, {Carey}, {Cudworth}, {Evans}, {Evans}, {Gates}, {G{\"a}nsicke},
  {Gillespie}, {Gilmore}, {Nebot Gomez-Moran}, {Grebel}, {Greenwell}, {Gunn},
  {Jordan}, {Jordan}, {Harding}, {Harris}, {Hendry}, {Holder}, {Ivans},
  {Ivezi{\v c}}, {Jester}, {Johnson}, {Kent}, {Kleinman}, {Kniazev},
  {Krzesinski}, {Kron}, {Kuropatkin}, {Lebedeva}, {Lee}, {French Leger},
  {L{\'e}pine}, {Levine}, {Lin}, {Long}, {Loomis}, {Lupton}, {Malanushenko},
  {Malanushenko}, {Margon}, {Martinez-Delgado}, {McGehee}, {Monet}, {Morrison},
  {Munn}, {Neilsen}, {Nitta}, {Norris}, {Oravetz}, {Owen}, {Padmanabhan},
  {Pan}, {Peterson}, {Pier}, {Platson}, {Re Fiorentin}, {Richards}, {Rix},
  {Schlegel}, {Schneider}, {Schreiber}, {Schwope}, {Sibley}, {Simmons},
  {Snedden}, {Allyn Smith}, {Stark}, {Stauffer}, {Steinmetz}, {Stoughton},
  {SubbaRao}, {Szalay}, {Szkody}, {Thakar}, {Thirupathi}, {Tucker}, {Uomoto},
  {Vanden Berk}, {Vidrih}, {Wadadekar}, {Watters}, {Wilhelm}, {Wyse}, {Yarger},
  \& {Zucker}}]{2009AJ....137.4377Y}
{Yanny}, B., {Rockosi}, C., {Newberg}, H.~J., {et~al.} 2009, \aj, 137, 4377

\bibitem[{{Zolotov} {et~al.}(2009){Zolotov}, {Willman}, {Brooks}, {Governato},
  {Brook}, {Hogg}, {Quinn}, \& {Stinson}}]{2009ApJ...702.1058Z}
{Zolotov}, A., {Willman}, B., {Brooks}, A.~M., {et~al.} 2009, \apj, 702, 1058

\end{thebibliography}

\end{document}